\documentclass[letter,fleqn,usenatbib]{mnras}   %instead of a4paper

\usepackage{newtxtext,newtxmath}
\usepackage[T1]{fontenc}
\usepackage{ae,aecompl}

\usepackage{graphicx}	% Including figure files
\usepackage{amsmath}	% Advanced maths commands
\usepackage{amssymb}	% Extra maths symbols
\usepackage{lscape}
\usepackage{CJK}

\def \micron{$\mu$m}

\def \msun {\rm M_{\odot}}

\def  \lx     {$L_{\rm X}$}
\def  \lagn     {$L_{\rm AGN}$}
\def  \lir     {$L_{\rm IR}$}
\def  \lfir     {$L_{\rm FIR}$}
\def  \lirsf     {$L_{\rm IR, SF}$}
\def  \lfirsf     {$L_{\rm FIR, SF}$}
\def \nh {$\rm N_{\rm H}$}

\def\sf{star formation}
\def \mbh {\rm M_{\bullet}}

\def  \ergs        {\hbox{erg s$^{-1}$}}              % erg/sec
\def  \kms         {\hbox{km s$^{-1}$}}          % kilometers per sec
\def  \lax    {${_<\atop^{\sim}}$}

\newcommand{\halpha}{\rm H{$\alpha$}}
\newcommand{\hbeta}{\rm H{$\beta$}}
\newcommand{\civ}{\rm CIV}
\newcommand{\mgii}{\rm MgII} %Mg {\rm II}

\begin{document}
\begin{CJK}{GB}{gbsn}
%%%%if fatal error in compiling, try \mbox{\citep{}}

% Please keep new commands to a minimum, and use \newcommand not \def to avoid
% overwriting existing commands. Example:
%\newcommand{\pcm}{\,cm$^{-2}$}	% per cm-squared

%%%%%%%%%%%%%%%%%%%%%%%%%%%%%%%%%%%%%%%%%%%%%%%%%%

%%%%%%%%%%%%%%%%%%% TITLE PAGE %%%%%%%%%%%%%%%%%%%

% Title of the paper, and the short title which is used in the headers.
% Keep the title short and informative.
\title[BHAR vs SFR in IR-bright AGNs]{Is there a relationship between AGN and star formation in IR-bright AGNs?}

% The list of authors, and the short list which is used in the headers.
% If you need two or more lines of authors, add an extra line using \newauthor
\author[Y. Sophia Dai et al.]
{\parbox{\textwidth}
{Y. Sophia Dai (´÷êÅ)$^{1,2,3}$\thanks{E-mail: daysophia$@$gmail.com},
Belinda J. Wilkes$^{4}$,
Jacqueline Bergeron$^{5,6}$,
Joanna Kuraszkiewicz$^{4}$,
Alain Omont$^{5,6}$,
Adam Atanas$^{7}$,
and Harry I. Teplitz$^{2}$}
\\
% List of institutions
$^{1}$Chinese Academy of Sciences South America Center for Astronomy (CASSACA), 20A Datun Road, Beijing, 100012, China \\
$^{2}$Caltech-Infrared Processing and Analysis Center, 1200 East California Boulevard, Pasadena, CA 91125, USA\\
$^{3}$UCLA, Department of Physics and Astronomy, Los Angeles, CA 90095, USA\\
$^{4}$Harvard-Smithsonian Center for Astrophysics, 60 Garden Street, Cambridge, MA 02138, USA\\
$^{5}$CNRS, UMR7095, Institut d$'$Astrophysique de Paris, F-75014, Paris, France\\
$^{6}$UPMC Univ Paris 06, UMR7095, Institut d'Astrophysique de Paris, F-75014, Paris, France\\
$^{7}$Picower Institute for Learning and Memory, Cambridge MA 02139\\
}
%Chinese Academy of Sciences South America Center for Astronomy, China-Chile Joint Center for Astronomy,
%Camino El Observatorio #1515, Las Condes, Santiago, Chile

% These dates will be filled out by the publisher
\date{Accepted XXX. Received YYY; in original form ZZZ}

% Enter the current year, for the copyright statements etc.
\pubyear{2018}

% Don't change these lines
\label{firstpage}
\pagerange{\pageref{firstpage}--\pageref{lastpage}} 
\maketitle

\begin{abstract}
We report the relationship between the luminosities of active galactic nuclei (AGNs) 
and the rates of star formation (SF) for a sample of 323 far-infrared (FIR)-detected AGNs. 
This sample has a redshift range of 0.2 $< z <$ 2.5, 
and spans three orders of magnitude in luminosity, ${\rm L_{X} \sim\,10^{42-45}}$\ergs.
We find that in AGN hosts, 
the total IR luminosity (8-1000\,\micron ) 
has a significant AGN contribution (average$\sim$20\%),
and we suggest using the FIR luminosity (30-1000\,\micron)
as a more reliable star formation rate (SFR) estimator. 
We also conclude that monochromatic luminosities at 60 and 100\,$\mu$\,m
are also good SFR indicators with negligible AGN contributions,
and are less sensitive than integrated infrared luminosities
to the shape of the  AGN SED, 
which is uncertain at $\lambda>$100\,\micron.
Significant bivariate \lx-\lir\ correlations are found, 
which remain significant in the combined sample
when using residual partial correlation analysis 
to account for the inherent redshift dependence. 
No redshift or mass dependence 
is found for the ratio 
between SFR and black hole accretion rate (BHAR),  
which has a mean and scatter of log (SFR/BHAR) $=3.1\,\pm$ 0.5,
agreeing with the local mass ratio between supermassive black hole and host galaxies.  
The large scatter in this ratio and the strong AGN-SF correlation found in these IR-bright AGNs
are consistent with the scenario of an AGN-SF dependence on a common gas supply,
regardless of the evolutionary model.
\end{abstract}


%%%%%%%%%%%%%%%%%%%%%%%%%%%%%%%%%%%%%%%%%%%%%%%%%%

%%%%%%%%%%%%%%%%% BODY OF PAPER %%%%%%%%%%%%%%%%%%

\section{Introduction}
\label{sec:intro}
One of the outstanding questions in the study of galaxy formation and evolution is 
how the presence of a supermassive black hole (SMBH)
influences the formation and physical characteristics of the host galaxy. 
A general connection has been confirmed both locally and at high redshift 
using empirical correlations between the SMBH mass ($\mbh$)
and the luminosity, mass, and stellar velocity dispersion of the host
 \citep[e.g.][]{kormendy95, ferrarese00,merloni10}.
A constant ratio has been found between $\mbh$ and the bulge mass ($M_{\rm bulge}$),
measured by several studies to be log($M_{\rm bulge}/\mbh)\sim$ 2.9 $\pm$ 0.5 \citep{magorrian98,mnf01,mnd02,mnh03},
or log($M_{\rm bulge}/\mbh)\sim$ 2.3 $\pm$ 0.3,
after correcting the $\mbh$ values by galaxy types \citep{knh13}. 
The scatter of this ratio
is found to increase at lower masses ($<10^{9.5}\,\msun$),
and is much larger in bulgeless or pseudobulge galaxies than 
in classical bulge or giant elliptical galaxies,
sometimes resulting in no observed correlations for the former types \citep[for review, see][]{knh13}. 
The general galaxy-BH coevolution picture is regardless
supported locally by the tight mass correlations,
and by the similar cosmic evolution of 
total \sf\ rate (SFR) and BH accretion rates (BHARs) up to $z =$ 3 \citep{silverman08, mnd14}.

Despite tremendous progress of the demographic studies of SMBHs, 
it is still debatable whether, and if so how, the SMBH regulates
the host galaxy formation. 
Various scenarios exist, sometimes resulting in opposite predictions.
The `feedback' process has been suggested by theories and simulations,
in which active BH accretion will suppress and eventually shut down 
star formation by heating or expelling the cold gas in the host
\citep[e.g.][]{snr98, dimatteo05,bower06, hopkins06, debuhr12, fabian12}.
In the merger-driven model, for example,
simulations predict that the merging of two galaxies will boost star formation and BH growth, 
until the feedback from active galactic nucleus (AGN),
quenches the SF,
especially from luminous AGNs \citep[e.g.][]{hopkins06, dimatteo05, treister12}.
Besides the merger model,
in several competing theories the BH and galaxy grow in tandem 
via accretion and SF,
especially for less-luminous AGNs \citep[e.g.][]{springel05, dekel09, ktp11, fanidakis12}.
Secular processes, 
internal to the galaxy, 
may lead to concurrent galaxy and BH growths 
based on a common gas supply, at lower galaxy mass \citep[e.g.][]{springel05}. 
Steady cold gas flow along cosmic filaments or quasi-hydrostatic dark matter halos
can contribute to the in situ BH accretion and star formation, 
but 
energy feedback from AGNs or supernovae is often needed to regulate
this process \citep[e.g.][]{lilly13, lapi14, aversa15, mancuso16}.
Despite all the simulation progress,
which shows that cosmic cold gas flows likely contribute to galaxy growth and eventually the BH growth, 
the mechanism whereby this material reaches or is ejected from
the center is not yet fully understood \citep[e.g.][]{bournaud11, gabornbournaud13, naabostriker17}.
In all scenarios, 
the connection between the central AGN and the star formation
is a key parameter to characterize the different models. 

Recent studies have tried to directly trace the global properties 
of AGNs and their host galaxies 
via correlations between their intrinsic luminosities, and their star formation rates (e.g. BHAR and SFR). 
Given the differences in spatial scales between AGN ($\sim$100\,pc) 
and \sf\ (up to tens of kpc),
any observed correlation would indicate an intrinsic connection \citep{alexhick12, knh13}. 
The AGN luminosities are often traced by X-ray luminosities (\lx) 
or optical emission lines.
Optical indicators (e.g. [OIII] and [OI]) 
are generally limited to narrow line regions with good spectral coverage,
and thus dominated by local type 2 AGNs (AGNs with signs of obscuration) 
or Seyferts \citep[e.g.][]{netzer09, diamondstanic12, matsuoka15}. 
The X-ray based AGN luminosities 
are generally more reliable than the bolometric luminosity based on optical continuum,
as the latter may still suffer from obscuration.
As a result, 
the X-ray luminosity, which is dominated by nuclear emission,
is more widely used as an AGN indicator,
where AGNs are commonly defined as systems with $L_X > {\rm 10^{42}\,\ergs}$
 \citep[e.g.][]{szokoly04, hasinger08}. 
As will be discussed below,
two outstanding factors affect the observed AGN-SF relations:
the method used to derive the SFR;
and the sample selection effects. 

\subsection{SFR Indicators}
Commonly-used SFR indicators
range from emission lines (e.g. \halpha),
ultraviolet (UV) luminosities, 
to luminosities in the mid-IR and total IR bands \citep[e.g.][]{kennicutt98}.
Unfortunately, 
all these methods have their weaknesses.
SFR inferred from optical emission lines are generally more reliable,
as they probe SFR on shorter timescales than the integrated UV or IR luminosities.
This method, however, suffers from limited sample size,
as emission line observations are time consuming,
especially for high-$z$ sources \citep[e.g. MOSDEF, ][]{mclean12}.
SFRs inferred from both UV and mid-IR luminosities
can be dominated or heavily contaminated by the AGN emission.
Without AGN removal, this results in overestimated SFRs 
as well as the total-IR luminosity
in these bands. 
In this study, for SFR indicator, we choose to use the far-IR luminosity.
The far-IR emission traces the cold dust
and provides a less contaminated measure of SFR in an AGN.
Besides integrated \lfir,
another common practice is to use a single FIR band luminosity as the SFR proxy (e.g. $Herschel$ PACS or SPIRE),
typically $\nu L_{\nu}$ (60\,\micron) \citep[e.g.][]{netzer09, shao10, santini12, rosario12, rosario13b},
or at longer wavelengths \citep[e.g., 90\,\micron, 100\,\micron][]{matsuoka15}. 
This is based on the assumption that at rest-frames greater than 50\micron,
the AGN contribution is insignificant.
However, the amount of the intrinsic AGN emission in the rest-frame FIR
remains uncertain at $\lambda >$ 40-50\micron\ \citep{dai12,podigachoski15}. 
For instance, 
by comparing different AGN SED templates from 
\citet{elvis94, richards06, netzer07, mullaney11, dai12, dale14},
we found an intrinsic variation of up to 0.9 dex at 60\,\micron\ 
between different AGN models (normalized at 6\,\micron).
Observationally, 
recent studies on local AGNs ($z < $ 0.05) reported
a FIR flux excess, possibly associated with AGN activity
\citep[e.g.][]{shimizu16}.
This motivates the use of the full SED
to deconvolve the AGN and SF contributions in several recent studies 
\citep[e.g.][]{chen15, stanley15, shimizu17, azadi17}
as well as in this work (Sec.~\ref{sec:sed}). 
Even studies using (far-) IR-based SFR report different, 
sometimes contradictory correlations (or lack of). 
The situation remains far from clear.

\subsection{Sample Selection Methods and Notes on Scatter}
AGN samples are typically selected in 3 bands:
X-ray, optical, and infrared \citep{padovani17}.
The majority of the AGN-SF correlation studies utilize X-ray selected AGN samples,
which are then matched to IR or sub-mm data.
Earlier studies based on X-ray and single band sub-mm detections 
found a luminosity dependent AGN-SF relation up to $z\sim$ 2.5:
significant correlation between \lx\ and SFR from $\nu\,L_\nu (60)$\footnote{Except \citet{lutz10}, where
SED based on 870\,$\mu$m observation was used to derive the SFR.}
in the most luminous (\lx\,$>10^{44}\, \ergs$) AGNs;
but no correlation at lower luminosities (or $z>$ 1) \citep{lutz10, shao10, rosario12, santini12}.
Common interpretations of these results
invoke different mechanisms at high and low AGN luminosities:
major mergers dominate the luminous end, 
triggering simultaneous BH accretion and starburst episodes;
while secular evolution is responsible for the growth of the majority of galaxies
with moderate nuclear activity. 
In the latter, non-merger driven \sf\ occurs
in step with SMBH accretion, possibly fueled by the same gas reservoir,
regardless of AGN activity, BH/host mass, or the level of obscuration \citep[e.g.][]{lutz10, rosario12, mullaney12b}.
Evidence of coeval AGN-SF evolution has also been found in 
massive galaxies, regardless of the level of SMBH accretion \citep{podigachoski15},
and in AGN samples of X-ray and FIR detections and SED based SFR \citep{xu15b}.
Similar luminosity dependent relations have also been observed 
between the AGN subtracted specific SFR (sSFR = SFR divided by stellar mass, ${M_*}$) and \lx,
where no correlation was found at \lx\ $< 10^{43-43.5}\, \ergs$ and $z < 1$ \citep{rovilos12, santini12}.
Stacking of the IR-undetected AGNs is a common practice.
It is worth noting that the majority of the stacked results are similar,
with either a luminosity dependent correlation that flattens towards the less luminous end \citep{lutz10, shao10, rosario12}
and lower specific \lx\,\citep[\lx/$M_*$][]{bernhard16},
or no overall correlations \citep{harrison12, mullaney12b, stanley15}.
Similar flatter correlation, or no SFR enhancement compared to regular main-sequence galaxies,
are observed in AGNs detected in both X-ray and FIR,
both locally \citep[$z<$ 0.05,][]{shimizu17} and at 0.5 $< z <$ 2.5 \citep{mullaney12b}.

In contrast, some studies have found that
X-ray selected AGNs show a strong negative 
relation between \lx\ and far-IR flux/luminosity.
This can be interpreted as suppressed host \sf\ from AGN feedback \citep{page12, barger15, shimizu15}.
Analysis of a larger sample showed that the \citet{page12} result was biased by the
limited sample size and cosmic variance \citep{harrison12}.
Nevertheless, more recently, suppressed \sf, possibly due to AGN feedback, has been reported,
either in the form of declining flux at 850 \micron\ towards higher \lx\ in quasars at $z>$1 \citep{barger15},
or with AGNs, mostly Seyferts and low-ionisation narrow-line emission radio galaxies (LINERs), 
lying below the main-sequence galaxies at $z<$ 0.05 \citep{shimizu15}.

Intrinsic X-ray obscuration makes the situation more complicated.
The obscuration is due to one or several of the following factors:
1. orientation-dependent obscuration
related to the central disk/torus-like geometry \citep[e.g. unification model,][]{barthel89,anm85};
2. other nuclear material, such as
the narrow emission line region, lying in a $\sim$ polar orientation; 
3. material along the line-of-sight through the host galaxy \citep{goulding12}.
Obscuration decreases the X-ray emission at soft energies, reducing the observed
flux, and thus X-ray-selected AGN samples retain a bias against obscured
sources. 
Observationally, several studies have noticed a lack of correlation
between (s)SFR and the obscuration (${\rm N_H}$)  levels \citep{lutz10,shao10,rovilos12,rosario12}.
Low levels of obscuration (\nh \lax $10^{23}$ cm$^{-2}$)
can be estimated from the observed hardness ratio (HR) for sources with
known redshift. However, as the obscuration of the
primary X-ray power-law component increases,
weaker soft X-ray components dominate the emission so that the HR
no longer traces the level of obscuration \citep{wilkes13}.
Low-frequency radio (e.g. 3CR) and high-energy X-ray samples, which
have little/no orientation bias, find that $\sim 50$\% of active galaxies
are obscured with $\sim 50$\% of these being
Compton thick (\nh\,$\sim\,10^{24-26}\, \rm cm^{-2}$, \citet{wilkes13,lansbury15,brightman16,lansbury17}).
The `observed' X-ray luminosities of high-redshift (z$\sim 1-2$),
high-luminosity 3CR sources are $\sim 100-1000 \times$ lower
than their unobscured counterparts for the most highly obscured ($\sim$ Compton
Thick, \nh \lax$ 10^{24}$ cm$^{-2}$) sources \citep{wilkes13},
and a subset will fall below the flux limit and be lost from the sample
altogether. For those that remain in the sample, hardness ratios
underestimate the obscuration levels, and thus the intrinsic
X-ray luminosities for $\sim 25-50$\% are also underestimated
by 1-3\,dex. This effect increases towards lower redshift
as the observed band moves towards lower energy.
Without accounting for these uncertainties, it is difficult to draw
conclusions on the presence/not of a relation between X-ray-based AGN
luminosities and SF.

For optically selected AGNs, on the other hand, 
different correlations have also been observed.
Positive correlations between \lagn, traced by [OIII] and [OI] lines,
and single band far-IR luminosity (60, 90, 100\micron),
have been observed in local ($z<$ 0.2), type 2 AGNs \citep{netzer09, matsuoka15}.
For broad-line, optical type 1 quasars, 
\citet{rosario13b} noticed an overall lack of 60\micron-based SFR enhancement 
in AGN hosts at 0.3 $< z <$ 2.1,
but recent studies found that this might vary with the level of star-formation.
Up to $z\sim$ 3, 
SFR increases with increasing optical-based \lagn, [CIV] line-width, and SMBH mass 
for moderate star-forming AGNs \citep[SFR $\sim\,300\,\msun\,yr^{-1}$,][]{harris16},
but remains constant in starburst AGNs (SFR $> 1000\, \msun\,yr^{-1}$) with 
higher \lagn, Eddington ratio, and SMBH mass \citep{pitchford16}.

Finally, selecting AGNs from star-forming galaxies,
i.e. by IR flux or luminosity,
has resulted in mainly positive correlations
between BHAR and SFR regardless of AGN luminosity ($L_{\rm AGN} = 10^{43-47}\,\ergs$).
This correlation exists in both X-ray selected star-forming AGNs with 
or without stacking the non-detections \citep{symeo11, chen13},
and in optical- or IR-selected star-forming AGNs \citep{chen15}.
A positive correlation suggests two possible scenarios of AGN/SF coevolution:
either a strong cold gas inflow is fueling the black hole accretion and galaxy \sf\ simultaneously, 
or a merger-triggered nuclear starburst with strong accretion
during the early encounter \citep{hopkins12}. 
Recent work combining X-ray, optical and IR AGN selections
did not find an AGN-SF (or BHAR-SFR) correlation,
and observed SFR bias by the AGN selections, 
with IR AGNs being more star-forming than optical AGNs,
and no SFR preference in X-ray selected AGNs \citep{azadi17}.

Regardless of how AGN samples were selected,
studies of various galaxy populations have found
AGNs lying mostly along the main 
sequence (MS) of star forming galaxies, within a relatively narrow range in
the ratio of SFR to M${_*}$ \citep[e.g.][]{noeske07, elbaz07, elbaz11, panella09, rodighiero11, speagle14, stanley17}.
Despite the general increase in SFR of the star-forming galaxy MS towards higher redshifts,
AGNs reside mainly in MS hosts exhibiting SFR and stellar mass similar to those of inactive star forming galaxies in 0 $< z <$ 3.
A small fraction ($<$10\%) of AGNs show enhanced average host SFR
\citep[e.g.][]{santini12, mullaney12a, rovilos12, rosario13a, rosario13b}. 
A positive correlation has been found 
between \lagn\ and circumnuclear SFR in local Seyfert galaxies \citep{diamondstanic12, esquej14, garciagonzalez16}.
Recent studies on long term BHAR indicates that
the apparent parallel growth observed for BHs and host galaxies may be 
primarily due to a joint dependence on stellar mass, 
in that the average SFR and BHAR are both larger in higher-mass galaxies \citep{yang17}.

Simulations show that a `real' AGN-SF correlation may be masked by the large scatter, 
possibly affected by various factors:
the AGN evolutionary stage of the sample included 
(e.g. for major mergers, the relation may differ before/during/after merging),  
the variability timescales of AGNs and SFR;
and the Eddington ratio (ER) distributions in the samples \citep[e.g.][]{hickox14, volonteri15b, stanley15}.
For example, a flat or non-correlation becomes significant and positive
when average instead of instantaneous \lx\ is used \citep{azadi15}; 
while the inclusion of upper limits or stacking may flatten the observed trend \citep[e.g.][]{stanley15}.
It is important to bear in mind that not all IR-bright galaxies are AGNs,
e.g. only 10-30\% of the (ultra-) luminous IR galaxies---(U)LIRGs---are AGNs \citep[e.g.][] {fu10, hopkins12}, 
and vice versa, 
not all AGNs are IR-bright.  
Moreover, the different ways of projecting the 
correlations may also affect the outcome. 
Data points are often binned to overcome poor statistics in assessing the trends, 
but this binning can introduce its own biases in the results. 
For example, ${\rm L_X}$ and SFR are not as strongly correlated when binned by AGN luminosity or BHAR
as when binned by SFR---a result that can be explained 
by the shorter timescales of AGN variability \citep[e.g.][]{gabornbournaud13, hickox14, chen15,volonteri15a}. 
Besides, as mentioned earlier, the way of measuring SFRs could also introduce systematics.

\subsection{This Paper}
The aim of this paper is to test the different galaxy evolution scenarios via the AGN-star formation connection 
with a statistically significant sample of active galaxies, 
undergoing both active AGN and star forming activities.
Since (i) X-ray surveys are typically dominated by AGN down to
L$_{\rm x}$ \lax 6 $\times 10^{-18}$ erg cm$^{-2}$ s$^{-1}$ \citep{luo17},
and are less biased against edge-on/obscured sources than optical surveys, 
and (ii) using full SED including FIR data
yields a more reliable SFR estimate for AGN systems,
in this work we choose to focus on IR-bright, X-ray selected AGNs
that are detected in both the X-ray and FIR. 
This sample is selected from the 11\,${\rm deg^2}$ X-ray Multi-Mirror Mission (XMM)-Newton 
Large Scale Structure (XMM-LSS) field, and all of them have known redshifts. 
In Section \ref{sec:sample}, we describe the multi-wavelength data and the AGN selection;
in Section \ref{sec:analysis} we calculate the \lir, SFR, SMBH mass, and Eddington ratios;
we then discuss our results and their implications in Section \ref{sec:results},
followed by a summary in Section~\ref{sec:summary}. 
In this work, we assume a concordance cosmology with $H_0=$ 70\,km\,s$^{-1}$\,Mpc$^{-1}$, $\Omega_{\rm M} =$ 0.3, and
$\Omega_{\Lambda}=$ 0.7.  

\section{THE SAMPLE}
\label{sec:sample}
To focus on the AGN phase
where both BH accretion and star formation are active,
we selected a sample of AGNs detected in both 
hard X-ray (2-10\,keV) and FIR (250\,\micron)
with redshifts and multi-wavelength photometry for SED and luminosity estimates. 

We started with the 10\,ks XMM-LSS X-ray deep full exposure catalog \citep[XLSSd,][C13]{pierre07,chiappetti13}.
The nominal flux limits (50\% detection probability) are
$3\,\times\,10^{-15}\,\ergs$ for the soft band (0.5-2\,keV), 
and $1\,\times\,10^{-14}\,\ergs$ for the hard band (2-10\,keV) over the survey region.
We restricted our sample to the 2,399 hard X-ray detected objects with either spectroscopic and/or photometric redshift ({\bf Parent sample}),
which consist of 75\% of the 3,194 hard X-ray detected objects in the field. 
The remaining 795 objects have no $z$ information
due to the non-uniform multi-wavelength coverage of the field.
These objects with no redshift estimate share 
a similar X-ray flux distribution but are generally fainter in the optical and IR.
We did not limit our sample to optical point-sources,
as extended optical morphologies have also been reported to be common in IR-detected AGNs \citep{dai14}.
For 50\% (1,190) of the hard X-ray targets, spectroscopic redshifts (spec-$z$) are available 
from: \\
a. the SDSS-BOSS DR12\footnote{http://skyserver.sdss.org/dr12} catalog (943, within a matching radius of 6\,\arcsec),\\
b. various publications (229\footnote{not counting the 301 objects with SDSS spectra in part a,
of which the redshifts are consistent in $>$97\% of the cases, and the spec-$z$ from BOSS was used.},
for detailed reference list see \citet{melnyk13}, M13), \\ 
c. an MMT-Hectospec redshift survey 
based on 24$\mu$m priors (18, see survey selection described in \citet{dai14}).\\
In parentheses are the numbers of unique spectra in these catalogs.
The remaining 50\% (1,209) objects 
have photometric redshifts (photo-$z$) reported in M13.

We then matched the parent sample 
to the HerMES DR3 and DR2 catalogs\footnote{http://hedam.lam.fr/HerMES/}
\citep{roseboom10,roseboom12,oliver12, wang14}
and identified 382 AGNs with 250\,$\mu$m detections ($> 3\sigma$). 
The HerMES XMM-LSS SWIRE field covers 18.87 deg${^2}$ 
and has a 1$\sigma$ sensitivity of 5.6\,mJy (instrumental $+$ confusion noise) at 250\,$\mu$m \citep{wang14}. 
A matching radius of 10\,\arcsec, between the 6\,\arcsec\ PSF for XMM and the 18\,\arcsec\ PSF for Herschel-SPIRE1 (250$\mu$m),
was chosen to maximize the matching counts while minimizing random associations to be $<$1.5\%. 

The rest frame, hard-band X-ray
luminosity (derived from 2-10 keV, hereafter referred to as
L$_x$) was determined assuming a photon index, $\alpha_{\nu} =1.7$ and
\nh (Gal) $=$ 2.6 $\times 10^{20}$ cm$^{-2}$ (\citet{chiappetti13},C13).
X-ray hardness ratios: HR $=$ (H $-$ S)/(H $+$ S), where S is defined as the net counts in the
soft band, 0.5-2 keV, and H is the net counts in the hard band, 2-10\,keV,
were determined from the net counts from C13.
\lx\ was corrected for obscuration based on the observed HR
for each source detected in both bands, and
for which HR $<-0.5$.
In this step we assumed an intrinsic power law spectrum with the same $\alpha_{\nu} $ and
\nh (Gal) values above.
For the 50 obscured sources with no soft band detection,
X-ray lower limits on the absorption corrected luminosity
were determined by adopting a conservative upper limit to the soft
band count rate of 0.005 ct s$^{-1}$ (Table 6 in \citet{pierre07}).
As discussed in Section~\ref{sec:intro}, the corrected X-ray luminosities remain
a likely lower limit as we cannot rule out the presence of an additional,
soft component in these low signal-to-noise data.
In addition, neither the total counts nor individual background estimates
is available in C13, so we were unable to determine the statistical errors on the individual HR.
As shown in \citet[][see their Fig 5]{aird15}, the errors on the X-ray 
luminosity estimate for sources with obscuration levels $\geq$ 10$^{23.5}$\ergs 
can be $\sim$1 dex.
To estimate the maximum level of error on our luminosity calculations, 
we assumed that the top 10\% most luminous sources were unobscured
at given redshift bins, and used their average value as an upper limit for the intrinsic luminosity.
This was done with a redshift bin size of 0.4. 
We then calculated the \lx\ uncertainties for the
lower luminosity (obscured) sources individually, following the above assumption.
The median and deviation of the \lx\ uncertainty calculated this way 
are $\sim$ 0.6 $\pm$ 0.4\,dex.
These values provide a conservative estimate for the \lx\ errors. 
For a more realistic estimate, 
we adopted an effective \lx\ upper limit based on the observed 6\,\micron\ luminosity,
following Equation~(1) introduced in Sec~\ref{sec:sed},
which has an intrinsic 1\,$\sigma$ scatter of 0.3\,dex.
The majority of the \lx\ uncertainty estimated this way
has a median of 0.3\,dex,
with a 1\,$\sigma$ deviation of 0.4\,dex, 
and a maximum value of 1.7\,dex. 
After adding the intrinsic scatter from the IR-\lx\ conversion,
these values are comparable with the (0.6 $\pm$ 0.4)\,dex calculated above. 

\begin{figure} 
\begin{center}
\includegraphics[width=\columnwidth]{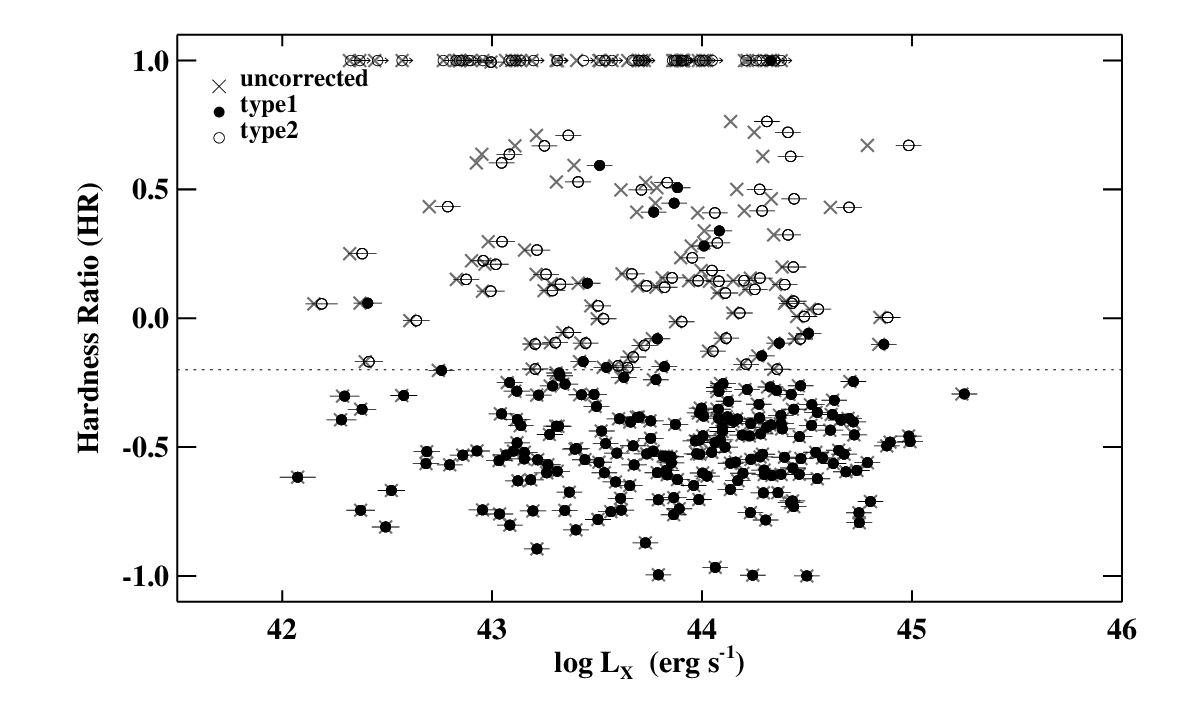}
\end{center}
\caption{The obscuration corrected (circles) and uncorrected (crosses)
\lx\ versus Hardness Ratio (HR, for definition, see Sec~\ref{sec:sample}) 
of the main sample. 
Filled circles mark the unobscured `type-1' AGNs (X-ray unobscured, HR $< $ - 0.2, or optical type 1), 
and open circles are the obscured `type-2' AGNs (X-ray obscured, HR $>$ -0.2).
About 14\% of the main sample has an X-ray lower limit. 
The dotted line marks the X-ray separation between type 1 and type 2 AGNs at HR $=$ -0.2 \citep{szokoly04}.
Above this dotted line, 
$\sim$15\% of the X-ray obscured sources show broad optical line features 
and qualify as type 1 objects.
\label{fig:hr}}
\end{figure}

Of the 382 AGNs with a $HerMES$ detection, 
328 are obscuration-corrected,
including 28 optical type 1 AGNs that are X-ray obscured. 
Most optical type 1 sources are unobscured (HR$<$-0.2, $\rm N_H < 10^{22}$ cm$^{-2}$ for z$\sim$2.5 and $\Gamma \sim 2$, \citet{hasinger08}).
A total of 166 (43\%) objects have HR$>$-0.2 and are defined as X-ray obscured (correction factor $>$1).
For the remaining 216 sources that are not X-ray obscured,
the \lx\ errors are $\sim 15-25$\% due to statistical errors combined with the
uncertainty in the spectral slope. 
Figure~\ref{fig:hr} shows the HR and $L_{\rm X}$ distribution of the main sample (defined below). 

In this study, we focus on the 323 far-IR detected X-ray AGNs
with ${\rm L_X \ge 10^{42} \ergs}$ and at $0.2 < z < 2.5$ ({\bf Main sample}, Table~\ref{tab:sample}).
Because of the requirement of FIR detection,
hereafter we will refer to this main sample as IR-bright AGNs.
This sample is reached after removing 26 sources with ${\rm L_X < 10^{42} \ergs}$
and 33 IR-bright AGNs outside this redshift range.
The redshift limits are motivated by the concerns
that
(1) at low $z$ ($z < $ 0.2 ) targets are more susceptible to obscuration
because the observed energy range is lower.
In fact, the absorption correction factor is on average $\sim$15\% higher 
below this $z$ cut.
In addition, our subsample at $z < 0.2$ 
has a $>$2$\,\times$ larger dynamical luminosity range than 
at higher redshift, with many sources below the $10^{42}$ \ergs\ cut;
(2) high $z$ targets are limited by small number statistics. 
The median and mean redshifts of the main sample are $z=$ 0.94 and 1.04, respectively.
About 60\% of the main sample have spec-$z$ (142 from BOSS, 29 from MMT, and 27 from M13), 
and the remaining 40\% are objects with photo-$z$ from M13. 
The multi-wavelength data associated with the X-ray sources 
are taken from the 2XLSSdOPT catalog (C13). 
A matching radius of 6\,\arcsec\ (PSF for XMM) is used
between the X-ray catalogs
and the GALEX, CFHTLS, SWIRE, and UKIDSS catalogs. 
Detailed description of the matching criteria
and references to the various catalogs can be found in C13.

\begin{figure*} 
\begin{center}
\includegraphics[ scale=0.42]{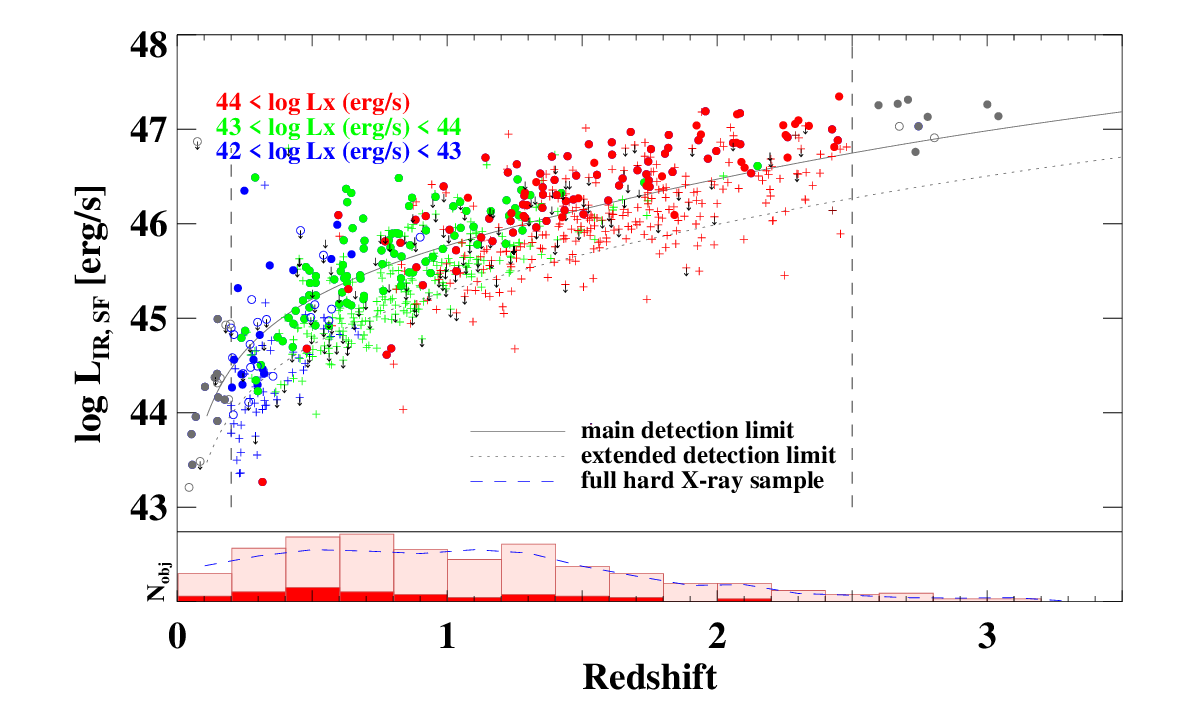}
\includegraphics[ scale=0.42]{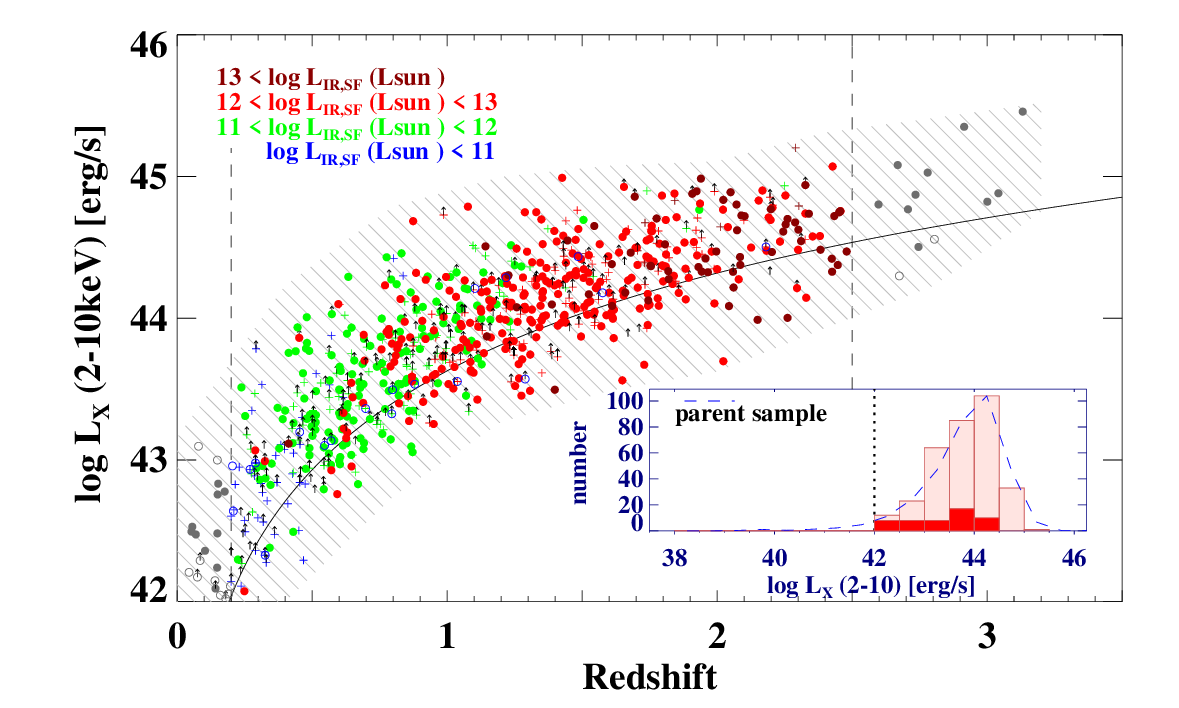}
\vspace{-10pt}
\end{center}
\caption{The AGN corrected infrared luminosity (\lirsf, {left})
and the absorption corrected X-ray luminosity (\lx, {right}) as a function of redshift.
Plotted are
the main sample of 323 IR-bright AGNs at $0.2 < z < 2.5$ (colored dots), 
the supplementary sample of IR-bright AGNs at $z < 0.2$ and $ z > 2.5$ (grey dots),
and the expanded sample of IR-undetected AGNs (crosses). 
The data points are color coded by the other luminosity, as labelled in the legend. 
Filled circles mark the unobscured AGNs (X-ray unobscured or optical type 1), 
which includes 65\% of the main sample.  
Open circles are the obscured AGNs (X-ray obscured and optical type 2). 
Arrows mark the X-ray obscured sources with an HR $>$ -0.2,
which are \lirsf\ upper limits (left) and \lx\ lower limits (right).
The black curves show the 3$\sigma$ (solid) and 1\,$\sigma$ (dotted) detection limits in the FIR (left)
and the nominal detection limit in the X-ray (right) (Sec.~\ref{sec:sample}).
Data points sometimes fall below the limits because of the AGN correction in the IR (Sec.~\ref{sec:sed}),
and the exposure time difference among X-ray pointings \citep{chiappetti13}.
The left inset shows the redshift distribution of the main sample,
and in red is the distribution for targets with \lirsf\ upper limits.
The right inset shows the distribution of 
the absorption corrected hard X-ray luminosity ${\rm L_X}$ (2-10\,keV) of the main sample,
and in red is the distribution for targets with X-ray lower limits.
\label{fig:zlx}}
\end{figure*}

The \lx\ in the main sample ranges from $10^{42.1}$ to $10^{45.5}\,\ergs$,
with a median of $10^{44.1} \ergs$ (Figure~\ref{fig:zlx}).
The majority (97\%) of the main sample has an \lx\ of $10^{42-45}\,\ergs$.
Half (166, 51\%) of the sample have an ${\rm L_X \ge 10^{44} \ergs}$;  
and the rest (158, 49\%) are at ${\rm 10^{42} \le L_X < 10^{44} \ergs}$,
confirming their AGN nature \citep[e.g.][]{szokoly04, hasinger08}. 
There are 46 obscured sources with HR $=$ 1,
whose reported \lx\ are lower limits.
An effective upper limit is given
for these objects, by applying a correction factor of $\sim$1.7\,dex,
which is the maximum \lx\ correction factor found 
for the rest of the sample.
About 60\% of the main sample has an HR $<$ -0.2 \citep[X-ray unobscured, e.g.][]{szokoly04}.
In the spec-$z$ subsample (198/323), 
55\% (109/198) show broad emission lines (optical type 1).
Combining both definitions,
overall 65\%\footnote{$\sim$15\% of the X-ray obscured sources show broad lines (optical type 1).} 
of the main sample are unobscured (Fig~\ref{fig:hr}).

For comparison purposes,
we retain the small subset of sources outside our preferred redshift range ({\bf Supplementary sample}, Table~\ref{tab:sample})
that satisfy the same luminosity and flux requirements,
to study the redshift and luminosity dependences.
The supplementary sample consists of 20 $z < 0.2$ objects
and 12 objects at $2.5 < z < 4.2$.

In addition, 
since a significant fraction (84\%) of the parent sample is not bright in the IR, 
we extend the IR limit to include fainter sources with `marginal' detections ({\bf expanded sample}, Table~\ref{tab:sample}).
The expanded sample includes 558 AGNs
with ${\rm L_X \ge 10^{42} \ergs}$ in 0.2 $< z < $ 2.5,
but formally undetected: their 250\,$\mu$m detection significance 
is between $1-3\,\sigma$.
The expanded sample will be used to characterize the effects of Malmquist bias 
commonly present in flux-limited samples.

In Table~\ref{tab:sample} we summarize the redshift and luminosity distributions for the 3 samples.
The basic physical properties including HR, 
the intrinsic N$_{\rm H}$, and the absorption corrected \lx\ 
are listed in Table~\ref{tab:data}. 
Figure~\ref{fig:zlx} plots the luminosities,  \lx\ and \lirsf\, 
as a function of redshift, color-coded by the other luminosity. 
The method to calculate the different IR luminosities is described in Section.\ref{sec:analysis}.

\subsection{Selection Effects}
\label{sec:bias}
As shown in Figure~\ref{fig:zlx} (right, inset),  
IR-bright AGNs in our main and supplementary samples
share similar $z$ and \lx\ distributions as the parent sample of 
hard X-ray detected targets (blue dashed line, scaled).
This indicates a limited influence on the intrinsic \lx\ distribution 
by the level of FIR activity.
Similar results have been found in radio AGNs \citep[e.g. 3C samples,][]{podigachoski15}, 
where the far-IR detection rate is unrelated to the radio source type (i.e. orientation). 

The redshift distribution, on the other hand,
shows a higher fraction of IR-bright AGNs at 0.2 $< z <$ 0.6 than in the parent sample.
This is mainly due to the IR detection requirement, 
as fainter objects at higher $z$ fall below the relatively shallow detection limit.  
The sharp drop of the number of objects at $z<$ 0.2 is due to the \lx\ lower limit of $10^{42.0}$ \ergs\,
and the small volume probed below this redshift,
which limit the number of luminous AGNs. 

In the main sample, 
the z $\sim$ 2 sources are systematically $\sim$ 1-2\,dex
more luminous in \lx\ than the z $\sim$ 0.5 objects (Figure~\ref{fig:zlx}),
due to the flux limit and larger volume probed at high redshift. 
This increase is broadly consistent with the increase in break luminosity ($L_{*}$) in the 
AGN luminosity function \citep[e.g.][]{croom09, ranalli16,aird15}.
This indicates that at all redshift in our selected range, 
we are sampling approximately the same portion of the AGN luminosity function relative to $L_{*}$.
Similarly, the SFR indicator \lirsf\ (for definition see Sec~\ref{sec:sed})
increases by 1-3 dex from $z=$ 0 to $z=$ 2 (Figure~\ref{fig:zlx}), 
comparable to the increase in SFR density 
and in the evolution of the typical ratio of SFR to 
stellar mass along the star-forming MS \citep[e.g.][]{speagle14}.
The expanded sample with marginal IR detections---formally undetected, 
is a continuation of the main sample to lower IR luminosities at all redshifts (crosses, Figure~\ref{fig:zlx}).
Inclusion of these 1-3\,$\sigma$ IR undetected sources 
provides information below the formal flux limit,
allowing us to check for systematic effects in the main sample due to Malmquist bias.

Finally, 
since  both spectroscopic and photometric redshift determination require optical spectra or photometry, 
the ER distribution is not homogeneous across redshift.
At high $z$, only high ER, luminous targets could be detected. 
We will discuss this specific selection effect in more detail in Section.~\ref{sec:er}.
These selection effects should be borne in mind when interpreting the results in Section~\ref{sec:results}.

\section{ANALYSIS}
\label{sec:analysis}
\subsection{IR luminosity, SFR, and Dust mass}
\label{sec:sed}
In this section we estimate the total IR and FIR luminosities (${\rm L_{IR}^{8-1000}, L_{FIR}^{30-1000}}$)
based on the rest-frame SEDs for the IR-bright AGNs.
The SEDs are constructed from optical through the FIR bands:
u*, g',r',i',z' (CFHTLS); 
J, H, K (UKIDSS); 
3.6, 4.5, 5.8, 8.0 \micron\,(SWIRE-IRAC);
24, 70, 160 \micron\,(SWIRE-MIPS); 
250, 350, 500 \micron\,(HerMES). 
For the Herschel data, the total errors (instrumental $+$ confusion noise) are
used in the fitting procedure.
We adopt the ${\rm T-\alpha-\beta}$ model from \citet{blain03},
where T is the dust temperature,
$\beta$ is the emissivity index,
and $\alpha$ the power-law index. 
This method fits the SED longwards of 5$\mu$m without any assumptions about the heating source,
be it AGN or star formation. 
Instead of a pure modified blackbody (MBB)
on both the Rayleigh-Jeans and Wien tails,
a power-law function ($f_{\rm \nu} \propto {\rm \nu}^{-\alpha} B(\nu,T_{\rm dust})$)
is used in the mid-IR (5-10\,\micron) Wien side to account for contributions from warmer dust. 
Here $B(\nu,T_{\rm dust})$ is the blackbody Planck function.
SED examples using the same method can be found in \citet{dai12}.
We adopt $\beta =$ 2.0 \citep{priddey03}
and allow $\alpha$ to vary.
This additional term is then matched to the MBB component at a transition point, 
where the two functions share equal zeroth and first order derivatives.
The transition wavelengths vary from case to case.
The corresponding peak dust temperature ranges from 5 to 100 K,
with a median around 30K, similar to normal star forming galaxies.
As a result of the larger errors in the FIR flux,
compared to the main and supplementary samples,
the expanded sample has a systematically 2-3\,$\times$ larger ($\sim$ 40-50\%)
uncertainty in their \lir\ and \lfir\ estimates.

Utilizing the X-ray data, we develop a 3-step method to 
decompose the AGN and star formation contributions in the FIR regime. 
Step 1 is to estimate the AGN contribution to the IR luminosity from the X-ray. 
This correlation is based on the assumption that the X-ray, especially in the hard band,
and mid-IR are both dominated by AGN emission. 
Here we choose 6\,\micron\ to enable extrapolation % instead of 12\,\micron\ 
into the far-IR regime because AGN SEDs may vary significantly longwards of
the rest-frame 10\,\micron\ for different AGN populations. 
For instance, in \citet{dai12} a variation on the order of 
1.5 dex was found between the 250\,\micron\ IR-detected and IR-undetected AGNs.
Several published relations exist regarding the X-ray to 6 \micron\ correlations
for AGNs with ${\rm L_X}$ in the range of $10^{41-46}\,\ergs$,
for both obscured and unobscured populations \citep[e.g.][]{lutz04, gandhi09, fiore09, lanzuisi09, mateos15, stern15, chen17}. 
In this work we adopt the results from \citet[]{stern15}:
\begin{equation}
log\,L(2-10\,{\rm keV}) = 40.981 + 1.024x - 0.047x^2
\end{equation}
where $L$(2-10\,keV) is in units of \ergs, 
and x $=$ log($\nu\,L_{\nu}(6\mu\,m)/10^{41}\ergs$).
This relation is consistent with earlier work at the fainter end
and covers a wide range of ${\rm L_X} = 10^{42-46}\,\ergs$,
which overlaps with the luminosity range of our sample. 

In step 2, we convert the X-ray based 6\,\micron\, luminosity (${\rm L_{6}}$) to the AGN IR (${\rm L_{IR, AGN}}$)
and bolometric luminosities (${\rm L_{AGN}}$) using an AGN template 
that extends to rest-frame 1000\,\micron\,  \citep[][D12]{dai12}. 
The D12 mean SED template is chosen because it was constructed
with detailed FIR SED information with SPIRE detections and stacks of FIR-undetected AGNs\,\footnote{https://app.box.com/v/dai12-templates},  
while earlier works, e.g. \citet[][R06]{richards06}, \citet[][N07]{netzer07}, \citet[][M11]{mullaney11}
stopped or extrapolated beyond rest-frame 100\,\micron\ 
where no data were available. 
Since the AGN contribution to the rest-frame FIR is an unsettled question
with a large variation ($\ge$1\,dex), 
in this study we adopt the D12 mean SED based on the stacks of 
$\sim$300 SPIRE-undetected AGNs.
This is likely an underestimate for the small subsample of 
AGN-starbursts ($\sim$10\% of all quasars according to D12),
whose \lir\ and \lfir\ are 0.3-0.4\,dex higher.
Given the redshift range used to construct the mean SED, 
the intrinsic uncertainty of this template increases from 
$\sim$0.3\,dex to $>$1\,dex beyond 100\micron.
Compared to the extrapolation of the above mentioned templates (R06, N07, M11, and \citet[][D14]{dale14}), 
the conversion factors between ${\rm L_6}$ and \lir\ are always consistent within 0.2\,dex.
However between ${\rm L_6}$ and \lfir,
the deviation is larger and varies from -0.06 (M11), 0.42 (R06), 0.47 (D14), to 0.53 (N07), respectively.
Regardless, these differences are 10 times smaller than the intrinsic scatter (covering 90\% of the sample) 
of a few dex and can be considered consistent with each other.
In summary, 
factors of 0.9 and 2.5 were used to convert ${\rm L_{6, AGN}}$ to ${\rm L_{FIR, AGN}}$ and ${\rm L_{IR, AGN}}$, respectively;
and a factor of 8.0 was used to convert the \lx\ based ${\rm L_{6}}$ 
to the AGN bolometric luminosity \lagn. 

\begin{figure*} 
\begin{center}
\includegraphics[ scale=0.6]{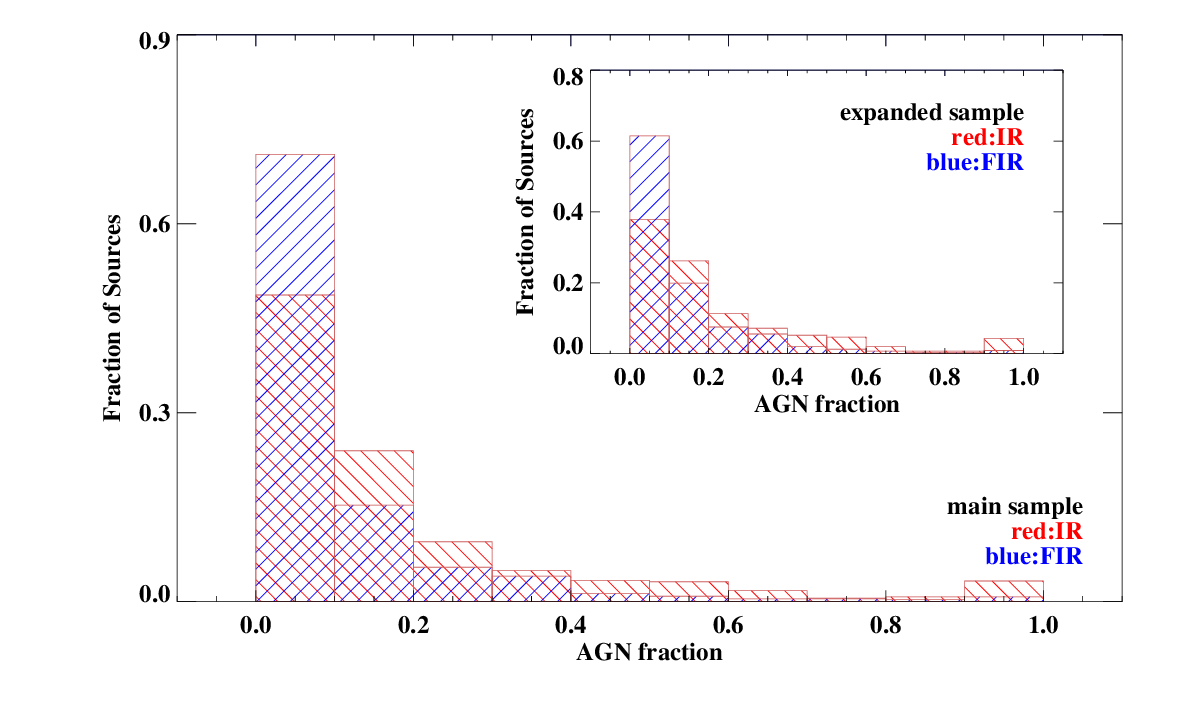}
\end{center}
\caption{The distribution of the fractional AGN contribution to the total IR (red) and FIR (blue) luminosities
for the main sample and the expanded sample (inset).  
The x-axis shows the $f_{\rm AGN}$ (i.e. $L_{\rm (F)IR, AGN}/L_{\rm (F)IR,obs}$),
and the y-axis marks the frequency of a certain $f_{\rm AGN}$.
For the main sample, 
AGN contributes an average of $>$11\% and 6\% to $L_{\rm IR}$ 
and $L_{\rm FIR}$, respectively. 
The actual percentage is higher than quoted here, 
since \lx\ in $\sim$15\% of the main sample are lower limits. 
In about 4\% (1\%) of the IR-bright AGN main sample, 
the IR (FIR) luminosity is purely AGN-heated 
(i.e. $L_{\rm (F)IR, AGN} \ge L_{\rm (F)IR, obs}$, thus SFR $=$ 0).
The reason for a smaller fraction of purely AGN heated \lfir\
than \lir\ is due to the AGN subtraction from the *observed* \lir\ and \lfir.
For the expanded sample, the average AGN contribution is 
at least 23\% (13\%) in the IR (FIR),
and 11\% (4\%) has a purely AGN-heated IR (FIR) luminosity. 
\label{fig:agnf}}
\end{figure*}

In the last step (step 3), we subtract ${\rm L_{IR, AGN}}$ and ${\rm L_{FIR, AGN}}$
from the observed ${\rm L_{IR}}$ and ${\rm L_{FIR}}$ derived from SED fitting,
and estimat the SFR based on the AGN-corrected \lirsf\ \& \lfirsf\
using the Kennicutt relation \citep{kennicutt98}\footnote{Note the definition of FIR in \citet{kennicutt98} 
equals the total IR (8--1000\,\micron). 
In this work, IR and FIR refer to ranges (8--1000\,\micron) and (30--1000\,\micron), respectively.}.
Figure~\ref{fig:agnf} shows the distribution of AGN contribution to the IR (FIR) in the main and expanded samples.
The average AGN contribution to the total IR luminosity (red) is at least 11\% in the main sample.
The actual percentage is higher than quoted here,
as \lx\ in $\sim$15\% of the main sample are lower limits.
More than $\sim$8\% has an AGN dominated ${\rm L_{IR}}$,
resulting in a $>50\%$ drop in the SFR,
and 4\% has a purely AGN heated ${\rm L_{IR}}$ (i.e. ${\rm L_{IR, AGN} > L_{\rm IR}}$, or SFR $=$ 0). 
As a result of the scatter in the $L_{\rm X}-L_{\rm 6, AGN}$ relation,
the uncertainties in \lirsf\ and SFR are also higher for objects with an AGN dominated IR.
On the other hand, for FIR luminosities (blue histograms in Figure~\ref{fig:agnf}), 
the AGN contribution is lower,
with an average value of at least 6\%, and only $\sim$1\% has an AGN dominated FIR.
The reason that \lfir\ has a smaller fraction of purely AGN heated sources 
than \lir\ is because of the different conversions from ${\rm L_{6, AGN}}$ to 
${\rm L_{IR, AGN}}$ and ${\rm L_{FIR, AGN}}$.
These uniform, template based conversions are 
subtracted from the observed SEDs,
which differ from the template on an individual basis.
As a result, the AGN subtracted
 \lirsf\ and \lfirsf\ values are not always correlated. 
For the expanded sample with lower IR luminosity, 
the fractional AGN contribution to the IR luminosities is higher, 
as expected given the constant X-ray flux limit (Figure~\ref{fig:agnf}, inset).
The average AGN contribution is at least 23\% in the IR,
and 26\% of the expanded sample has an AGN dominated IR ($>50$\% drop in the SFR),
and 11\% has purely AGN heated ${\rm L_{IR}}$ (SFR $=$ 0). 
For \lfir,
AGN contribution has an average of 13\%, 
and 4\% of the expanded sample has an AGN dominated FIR. 

These high values of correction factor demonstrate the importance of IR AGN/SF decomposition 
for SFR estimates.
It is worth noting that the average AGN fraction in the IR 
increases with redshift. This is a known selection effect due to 
converting the observed 250\,$\mu$m to the rest-frame with a fixed, and steep, 
SED template. 
This results in the inclusion of galaxies with
relatively lower \lfir\ at similar \lx,
as the observed frame approaches the IR SED peak.

We compare the AGN-removed ${\rm L_{IR, SF}}$ to
the total ${\rm L_{FIR}}$ and find that they are consistent within errors for 92\% of the main sample. 
Therefore we suggest that when AGN decomposition is not possible,
${\rm L_{FIR} (30-1000\,\mu m)}$ can be used as a convenient proxy
for the AGN removed ${\rm L_{IR, SF}}$. 
As a check, we also subtract the average contribution to ${\rm L_X}$ from star formation
using the SFR-$L_{\rm X}$ relation \citep{ranalli03},
and confirm that $L_{\rm X}$ is dominated by the AGN:
the non-AGN contribution to $L_{\rm X}$ is $<$ 2\% 
in all chosen redshift and luminosity bins. 

We then estimate the dust mass ($M_{\rm dust}$) of the sample
using the following formula \citep{beelen06}:
\begin{equation} \label{eqn:mdust}
M_{\rm dust} = \frac{S_{\nu 0}D_{\rm L}^2}{(1+z) k_{\rm d}(\nu) B(\nu,T_{\rm dust})}
\end{equation}
where $k_{\rm d}(\nu)=k_0(\nu/\nu_0)^\beta$ is the dust absorption coefficient,
$T_{\rm dust}$ and B are the dust temperature 
and the black body Planck function,
and $D_{\rm L}$ is the luminosity distance based on redshift. 
Here we use the flux at 250$\mu$m $S_{\rm 250}$, and $k_{\rm d}$ from \citet{alton04}.
The majority (86\%) of the sample has ${\rm log}\,M_{\rm dust} > 10^8\,\msun$ (99\% at $> 10^7\,\msun$)
similar to the dust-rich quasars detected in the FIR and (sub)mm \citep[e.g.][]{dai12}.
This value is $\sim$1-2 dex higher than the dust mass estimated for the local Palomar-Green (PG) quasars,
confirming that this IR-bright AGN sample is dominated by objects with ample dust,
likely in the process of actively forming stars.
Table~\ref{tab:data} lists the derived properties of the sources for the IR-bright AGN samples (main, supplementary, and expanded).
The full table is available in a machine-readable form of the online journal.

\subsection {SMBH mass, Eddington Ratios, and BHAR}
\label{sec:er}
About 90\% of the optical type 1 (broad-emission-line) AGNs
in the main sample (i.e. 34\% of the full main sample)
have a spectrum of sufficiently high signal-to-noise
to derive reliable virial SMBH masses ($\mbh$).
Note that the AGN luminosities for targets detected only in the X-ray hard band
are lower limits. 
The virial SMBH masses are commonly expressed as \citep[e.g.][]{dai14}:
\begin{equation} \label{eqn:virial}
 {\rm log} \left(\frac{\mbh}{\msun}\right) = a + b\,{\rm log} \left(\frac{\lambda L_{\lambda}}{10^{44} \ergs}\right) + c\,{\rm log} \left(\frac{\rm FWHM}{\kms}\right)
\end{equation}
where $\msun$ is the solar mass, 
FWHM is the full-width-at-half-maximum of the emission line profile, 
and ${\rm \lambda L_{\lambda}}$ is the continuum luminosity at 
5100$\AA$ (\hbeta, \halpha), 3000$\AA$ (\mgii), and 1350 $\AA$ (\civ), respectively. 
The term $\lambda L_{\lambda}$ is used as a proxy for the radius of the broad line region \citep{kaspi00, bentz13}.
The coefficients $a$ and $b$ are empirical values 
based on SMBH masses determined via the reverberation mapping method,
and $c$ normally has a fixed value of 2 \citep[e.g.][]{vp06},
which exemplifies the virial nature of the broad line region ($\mbh \propto G v^2 R^{-1}$). 
Here we use the FWHM (in $\kms$) of the continuum subtracted emission line as the line width proxy.
We adopt the IDL line fitting procedures
from \citet[][Sec 3]{dai14}
for \civ\ (0.660, 0.53, 2.0), \mgii\ (0.740, 0.62, 2.0), \hbeta\ (0.672, 0.61, 2.0), 
and \halpha\ (0.522, 0.64, 2.06) lines;
in brackets are the parameter sets (a, b, c)  from \citet{vp06, shen11, md04, gh05}, respectively. 

\begin{figure*} 
\begin{center}
\includegraphics[ scale=0.6]{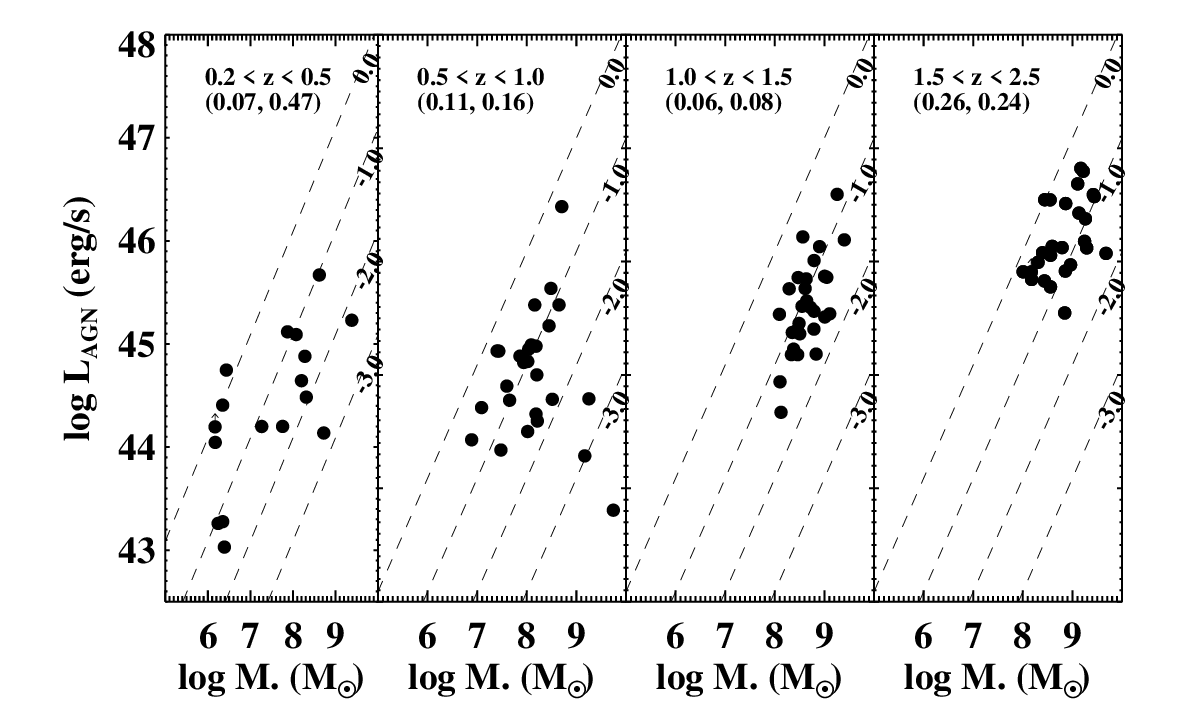}
\end{center}
\caption{The AGN bolometric luminosity vs. SMBH mass ($\mbh$) 
for the subsample of IR-bright AGNs with reliable BH mass estimates 
($\sim$34\% of the main sample).
Dashed lines mark the Eddington ratios (ERs)  at 1, 0.1, 0.01, and 0.001, and are labeled by log (ER). 
At $z < 1.0$, the IR-bright AGN sample has a wide range of ER: 0.001 $<$ ER $< 0.01$,
whereas at high $z$, the sample is limited to massive, ER $>$ 0.01 AGNs.
In brackets are the median and standard deviation of the ER in each redshift bin.
\label{fig:mbh}}
\end{figure*}

For the subsample with spectra showing broad-emission-lines of sufficient quality, 
we use the $\mbh$ 
and \lagn\ calculated in Sec~\ref{sec:sed},  
and compare the ER ($L_{\rm AGN}/L_{\rm edd}$) in four fiducial redshift bins (Figure~\ref{fig:mbh}),
where $L_{\rm edd} / \ergs = 1.3 \times\,10^{38} (\mbh / \msun)$.
The median ER shows a general increase from low $z$ to high $z$.
At high $z$, low mass AGNs are generally not detectable 
unless the ERs are sufficiently high that \lx\ is above the detection limit.
This ER selection effect is less prominent at $z <$ 1.5, 
where the ER distribution shows a wide range
and scatters into the ER $<$ 0.01 region.
Whereas from $z=$ 0.5 to $z =$ 1.5, 
more luminous AGNs are being selected,
the data points are distributed along constant ERs, 
indicating systems of similar accretion conditions.

We also calculate the $BHAR (\dot{M}_{\bullet}$) using the hard \lx\ as a proxy 
\begin{equation} \label{eqn:bhar}
\frac{BHAR}{\msun\,yr^{-1}}  = 0.15 \frac{0.1}{\epsilon}\frac{k L_{\rm X}}{10^{45}\ergs}
\end{equation}
where $\epsilon$ is the mass-energy conversion efficiency,
and $k$ is the conversion factor between \lx\ and the AGN bolometric luminosity. 
Here we adopt $k =$ 22.4 from \citet[][based on local AGNs]{vnf07},
and a typical $\epsilon$ value of 0.1 \citep{marconi04},
meaning that about 10\% of the mass is converted into radiative energy. 
These values are chosen to allow direct comparisons 
with other studies involving $\dot{M}_{\bullet}$ estimates \citep[e.g.][]{mullaney12b, chen13}.

\section{RESULTS AND DISCUSSION}
\label{sec:results}
\subsection{Correlation between AGN activity and Star Formation}
\label{sec:correlation}
\begin{figure*} 
\begin{center}
\includegraphics[scale=0.7]{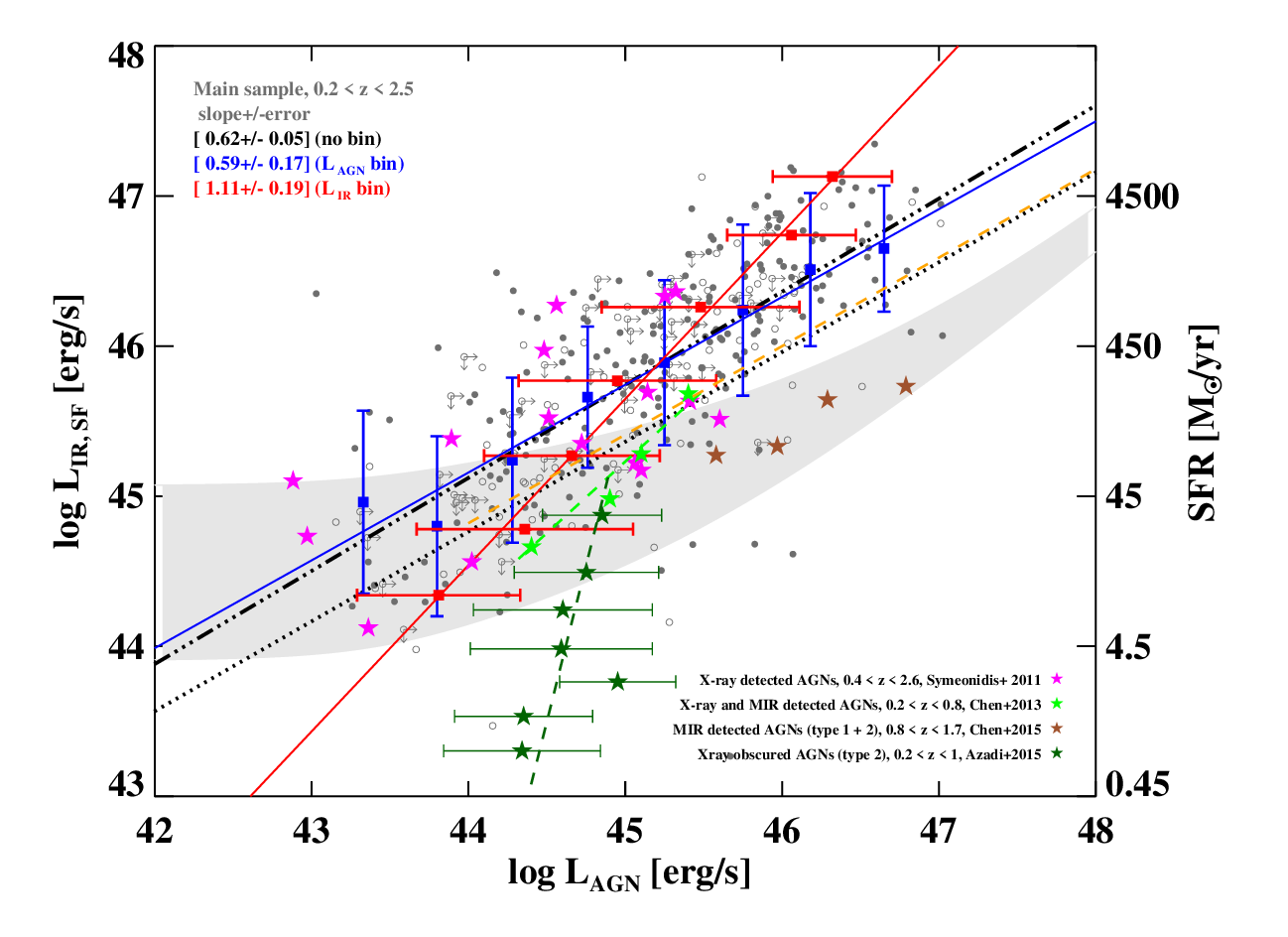}
\end{center}
\caption{
Correlation between ${\rm L_{AGN}}$ and 
${\rm L_{IR, SF}}$ for the main sample.  
Individual IR-bright AGNs are plotted as open and filled grey circles, same as in Figure~\ref{fig:zlx}. 
The best fit correlation (without binning) is marked by the dash-dotted black line,
and the same relation for the main$+$expanded$+$supplementary sample by the dotted black line. 
In blue is the average ${\rm L_{AGN}}$ plotted in bins of ${\rm L_{IR,SF}}$, 
in red is the average ${\rm L_{IR,SF}}$ plotted in bins of ${\rm L_{AGN}}$.
The shaded region marks the \citet{hickox14} model at 0.01 $< z < $ 3.5.
Also plotted are the binned data of IR-bright AGNs from the literature:
pink stars are the X-ray and FIR detected AGNs from \citet{symeo11},
green and brown stars are the X-ray and mid-IR detected AGNs from \citet{chen13,chen15},
the dark green stars are the X-ray obscured AGNs from \citet{azadi15}.
Dashed colored lines mark the 
literature correlations in \citet{chen13} (light green), \citet{azadi15} (dark green), and \citet{ xu15b} (orange), respectively.
Our results are in good agreement with  \citet{xu15b},
and in general agreement with \citet{chen13}.
The selection effects and caveats 
are discussed in Sec.~\ref{sec:results}.
\label{fig:lxsfrcf}}
\end{figure*}

In Figure~\ref{fig:lxsfrcf}, we compare the \lagn-\lirsf\ relation of the main sample
to literature results. 
Individual objects are plotted as grey dots (X-ray unobscured or optical type 1) 
or open circles (X-ray obscured and optical type 2).
The $L_{\rm X}$ lower limits and accordingly, \lirsf\ upper limits of the HR $=$ 1 sources are 
marked by arrows. 
The thick black dash-dotted line 
shows the best-fit correlation for the main sample, 
with a power-law of \lirsf\,$\propto\,$\lagn$^{(0.62\,\pm\,0.05)}$
and a high significance (P $<$ 0.0001).
We note that, since our sample is flux limited, 
the effective IR luminosity limit is higher at higher redshift. 
Thus the fainter end of the IR-bright AGN population is missed especially at high $z$,
also increasing the average \lir\ (Figure~\ref{fig:zlx}).
We will probe this fainter population later using the combined sample
 (main$+$expanded, See Section~\ref{sec:psra} \& \ref{sec:ratio}).  

Our best-fit correlation agrees well 
with the \citet[][X15]{xu15b} result (dashed orange line),
which shares almost identical selection criteria,
except an additional selection using the MIPS 24$\mu\,m$ flux.
The 24$\mu$m flux selection is
highly complete for AGN populations \mbox{\citep{krawczyk13,dai14}},   
as is further demonstrated by the similar correlations found between X15 and this work.
Our correlation is also in general agreement with 
\citet[][light green stars in Figure~\ref{fig:lxsfrcf}]{chen13},
after taking the scatter and binning methods into account.

The steeper \citet{azadi15} results were based on 
data binned by \lir\ (see also Sec.~\ref{sec:bin}).
The \citet{hickox14} model (grey shaded area) under-predicts 
the \lirsf\ for the most luminous IR-bright AGNs in our sample.
This model flattens at lower luminosities after accounting for the
effects of short-term AGN variability.
The disagreement between this work
and \citet[][, C15]{chen15} can be explained by the 
different sample compositions. 
C15 includes a high fraction of
type 2 AGNs (brown stars) as well as stacks of FIR-undetected objects,
thus their averaged values occupy a lower IR region than our correlation.

Another cause of the observed differences between our sample and some literature results
is the use of different SFR estimators.  
In Figure~\ref{fig:lxsfrcf}, 
the 15 AGN-ULIRG/LIRG systems from
\citet{symeo11} are systematically higher in \lir\ than our sample 
(with a $\sim$ 0.2\,dex IR offset at similar \lagn).
This is because their SED library used, \citet{siebenmorgen07},
was based on pure star-forming galaxies. 
As demonstrated earlier, the AGN contribution to \lir\ and \lfir,
though small (with an average of 23\% and 11\%),
is not negligible.
If we correct the \lir\ by applying the empirical \lir-to-\lfir\ correction
in our sample 
and use \lfir\ as a proxy for \lirsf\ (See sec~\ref{sec:sed}),
the \citet{symeo11} data agree better---the offset drops to $\sim$ 0.05\,dex.
Careful treatment of AGN removal is needed when attributing higher \lir\ to 
enhanced star formation in the AGN-host system.

The different SED models also bias the SFR estimates.
For instance, \citet{azadi15} 
utilized the iSEDfit code \citep{moustakas13},
which was based on the UV and optical photometry.
This code accounts only for unobscured star formation with no AGN removal. 
It yields systematically lower SFR estimates than 
using {\it Herschel} FIR data \citep[][ Section~\ref{sec:bin}]{azadi15}, 
as dust-reprocessed (IR)  extinction was not included.
This explains their systematically lower values in Figure~\ref{fig:lxsfrcf} (dark green stars).

Several earlier studies used 
a single band rest-frame FIR photometric measurement 
as the proxy for \sf.    
Since there is a non-negligible AGN contribution to the IR,
especially at $\lambda < 30\,\mu$m,
as demonstrated in Sec~\ref{sec:sed},
this approach may over-estimate the SFR.
Besides the \lir\ and \lfir\ defined earlier,
to test this at longer wavelength, 
we estimate the integrated luminosities at $L_{\rm 60-1000}$
and $L_{\rm 100-1000}$, based the same SED fitting procedure used in Sec~\ref{sec:sed}.
We find an average AGN contribution of $\sim$7\% in both luminosities.
If interpolated to single bands,
the AGN contribution to $L_{\rm 60}$ and $L_{\rm 100}$
is $\sim$2\%, almost negligible.
The higher AGN contributions to integrated values than at single bands
are due to the D12 AGN SED template being flatter towards longer wavelengths ($\lambda\,>$100\,\micron). 
Note that in our case any monochromatic luminosity
is interpolated from the SED fit and so is not independent
from the integrated values.
Actual AGN contributions may vary case by case at these wavelengths. 

Since Figure~\ref{fig:lxsfrcf} is plotted in luminosity space,
it is important to separate any real 
correlation from effects resulting from the presence of redshift on both axes.
The flatter than linear (1:1), 
$\alpha\sim0.6$ slope in Figure~\ref{fig:lxsfrcf} 
indicates a true AGN-SF correlation, 
but is also affected by a number of factors,
such as the increasing AGN fraction with redshift (see Section~\ref{sec:sed}), 
and the X-ray absorption correction, which broadens the range of \lx.
To test the validity of this correlation, 
we first compare the fit results across our various samples,
as discussed below.  
Then, in Section~\ref{sec:psra}, we use the partial correlation technique 
to examine the correlation between different parameters.
We will explore the binning effects in Section~\ref{sec:bin}.

Combining the main and expanded samples 
significantly increases, sometimes doubles, 
the IR luminosity parameter space at any given redshift.  
Fitting the combined sample results in a consistent slope 
of 0.60 $\pm$ 0.03 (dotted black line in Figure~\ref{fig:lxsfrcf}),  
with a lower normalization factor,
agreeing better with the \citet{chen15} results.
Similarly, including the supplementary sample at the low and high $z$ ends
results in a wider redshift range (0.04 $ < z < $ 4.2), 
but the slope remains consistent at 0.63 $\pm$ 0.04. 
The consistency of slopes estimated from fitting various subsamples 
confirms that the observed correlation is not purely caused by the 
Malmquist bias. 
To further test the effect of flux limits on the observed correlation,
we artificially increase the flux limits by factors of 5, 7, 9, and 11. 
Consistent slopes and normalization factors are found,
with slope values at 0.60\,$\pm$\,0.07, 0.51\,$\pm$\,0.11, 
0.59\,$\pm$\,0.18, and 0.58\,$\pm$\,0.23, respectively.
This confirms that the observed trend is intrinsic and not caused
by the IR flux limit of the sample. 

For the subsample of broad-emission-line AGNs with a reliable $\mbh$ estimate,
we also check the effects of ER and $\mbh$ on the \lagn-SFR relation
by binning the data by accretion efficiencies. 
Positive linear correlations are confirmed,
although the smaller sub-samples do not provide meaningful constraints on the slopes. 
We conclude that neither the mass nor the ER 
of the SMBH
regulates the AGN-SF correlation significantly, 
at least not on a timescale short enough to affect the 
observed star formation.

\begin{figure*} 
\begin{center}
\includegraphics[scale=0.6]{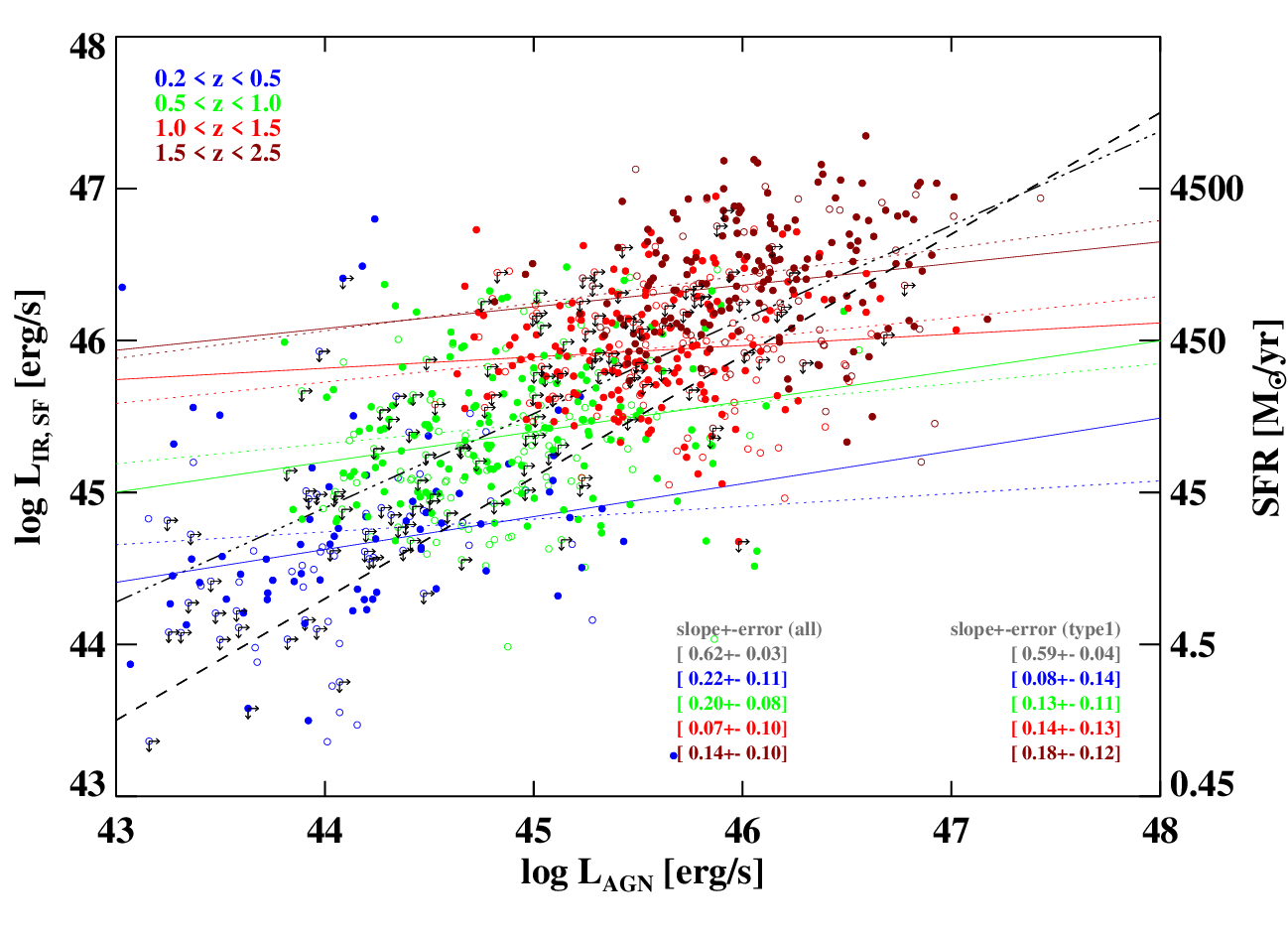}
\end{center}
\caption{
Correlation between \lagn\ and the AGN removed \lirsf\
for the main$+$expanded sample, color coded by redshift. 
The filled (open) circles are the same as in Figure~\ref{fig:zlx}.
Objects detected in the hard band only (X-ray lower limits) are marked with arrows.
The dash-dotted line is the correlation for the main sample without binning (See Figure~\ref{fig:lxsfrcf}).
The solid lines are the linear fits to the data in each redshift bin,
and the dotted lines are the fits for type 1 AGNs.
The dashed line marks the normalized
 \citet{netzer09} relation implied from local type 2 AGNs, with a slope of 0.8.
Significant correlations with flatter slopes are observed between \lagn\ and \lirsf\ in the limited redshift sub-samples.
\label{fig:lxsfr}}
\end{figure*}

\subsection{Correlation Analysis}
\label{sec:psra}
We then test for the presence and significance of correlations between the fluxes and
luminosities at various wavelengths across the SED.
Luminosity vs luminosity correlations are a challenge to assess
since both parameters depend on the redshift to
derive the L values from the observed fluxes.
Besides, both flux and luminosity correlations can be strongly affected by
selection effects for a given sample. 
In this section we take advantage of our large, well-defined sample 
to test the inter-relationships between these variables,
with an emphasis
on assessing the correlation significance independent of redshift.

Table~\ref{tab:bcamain} and Table~\ref{tab:bca} summarize the results
of the bivariate correlation analysis for the main and the
combined samples, respectively.  We test the
correlations between \lx\ vs L$_{\rm 60}$, \lx\ vs L$_{\rm 100}$,
\lx\ vs \lir, \lx\ vs \lfir, 
\lx\ vs \lirsf\, and \lx\ vs \lfirsf,
where L$_{\rm 60}$ and L$_{\rm 100}$
are the luminosities at 60~$\mu$m and 100~$\mu$m,
calculated from the SED fitting. 
A correlation is considered significant if the probability ($P$) 
of occurring by chance is $P < 0.01$.  
We find positive and significant bivariate correlations between the
AGN (\lx) and all IR luminosities.
However, in the main sample there is no bivariate correlation
between the rest-frame fluxes ($F_{\rm X}$ vs $F_{60}$ or $F_{\rm X}$ vs $F_{100}$;
Table~\ref{tab:bcamain}, and \ref{tab:bca}),
suggesting that their observed, strong  AGN-IR luminosity correlations
are primarily redshift driven. 
Adding sources below the flux limit in the combined
sample probes the fainter IR population, 
and includes information on the luminosity distribution by 
retaining
the individual estimates for each source. Bivariate correlation analysis
reveals a marginal correlation  for $F_{\rm X}$ vs $F_{60}$ ($P=$ 0.0106)
and a significant correlation  for $F_{\rm X}$ vs $F_{100}$ ($P=$ 0.0073),
suggesting that a residual AGN-SF correlation may be present.

We next perform
partial Spearman rank analysis \citep[PSRA,
e.g.][]{kns76,isobe86,ans96} between the AGN and IR properties. PSRA
allows for a correlation analysis in the general multivariate case,
and tests for correlations between subsamples of 
parameters while holding constant all other variables in the matrix.
In particular, these tests allow us to investigate correlations
independent of the, otherwise dominant, effect of redshift. 
To account for lower limits in our data\footnote{Absorption corrected 
X-ray flux and luminosity  with
HR$=$1 were treated as lower limits.}, we 
use the survival analysis package ASURV \citep{lavalley92} to
calculate the bivariate Spearman ranks that are then input to PSRA.

All luminosities (\lx, \lir, \lfir, L$_{60}$, L$_{100}$) are found to
be primarily and significantly correlated with redshift ($P<0.005$, not listed in the table). 
The partial correlation probabilities between pairs of luminosities and fluxes
are given in Table~\ref{tab:pca}. 
No significant correlations are found in the main sample
after removing the redshift effect,
except for \lx\ vs $L_{\rm IR}$. 
However, for the combined sample, significant correlations are present 
in  $F_{\rm X}$ vs $F_{\rm 100}$ and
$L_{\rm X}$ vs $L_{\rm 100}$, 
as well as in $F_{\rm X}$ vs $F_{\rm 60}$
and $L_{\rm X}$ vs $L_{\rm 60}$.
The partial correlations between \lx\ and the
broad-band IR luminosities (\lir, \lfir), 
are marginally significant ($P=0.011$, $P=0.061$, respectively).
Since our primary motivation is to determine whether or not there is 
a correlation between star formation and AGN, we also test
the partial correlation between \lirsf\ vs \lagn.
However, this correlation is not significant. 

Considering the results of all these correlations, 
along with the relatively low AGN contribution at these single bands 
($\sim$2\%, Sec~\ref{sec:correlation}), 
we used $L_{\rm 60}$  and $L_{\rm 100}$,
interpolated from the SEDs,
as reliable proxies for the SFR. 
The significant partial correlations between $L_{\rm X}$ vs $L_{\rm 100}$
and $F_{\rm X}$ vs $F_{\rm 100}$ thus suggest an AGN-SF connection, 
which remains significant after accounting for redshift dependence. 
The lack of correlation between \lirsf\ and \lagn\ is consistent 
with this conclusion if the uncertainties
and increased dispersion 
arising from the AGN subtraction process 
masks any real correlation, as suggested by simulation results \citep{gabornbournaud13, volonteri15a,hickox14}.
The main uncertainties in this sample include:
\lx\ obscuration correction (typically $\sim$ 0.6\,dex, 
up to 1.7\,dex for the $\sim$14\%, HR $=$ 1 sources, Section~\ref{sec:sample});
\lagn\ estimate (typically $\sim$\,0.3\,dex, up to 0.5\,dex, due to the \lx-$L_{\rm 6}$-\lagn\ conversion, 
and is AGN template dependent, Section~\ref{sec:sed});
\lirsf\ and thus the SFR estimate (typically $\sim$0.3-0.4\,dex, 
up to $>$1\,dex for outliers from the SED template, Section~\ref{sec:sed}).
After removing the redshift effect,
our sample shows a slope of $\sim$0.1-0.3 (see next paragraph), 
which translates to an IR luminosity increase of $\sim$0.4-1.2\,dex
over the \lagn\ span of $\sim$4\,dex in our sample. 
The typical combined error for \lx$+$SFR is $\sim$0.9\,dex ($\sim$1.2 if \lagn\ is used),
but for the extreme outliers ($<$ 3\%), it can reach an order of $\sim$2-3\,dex.
If combined with the intrinsic scatter,
this typical error is comparable to,
and thus sufficiently large to 
mask out the underlying intrinsic correlation.

An alternative method of removing the strong redshift effect 
in the bi-variate correlations is to test for a correlation over a smaller range of redshift. 
We compare the least square linear fit between \lagn\ and \lirsf\
for four sub-samples with redshift ranges:
0.2 $< z < $ 0.5, 0.5 $< z < $ 1.0, 1.0 $< z < $ 1.5, and $1.5 < z < 2.5$. 
Over the full redshift range, there is a strong overall bivariate
 \lx-\lirsf\ correlation ($P <$ 0.0001, Section~\ref{sec:correlation}).
For the main sample, 
within these smaller, sub-samples 
we find significant correlations in the intermediate redshift bins 
 ($P =$ 0.32, 0.006, 0.007, 
and 0.016, $z$ from low to high, no binning),
suggesting a weak ($\alpha \sim$ 0.2) underlying AGN-SF relation.
Similar to the partial analysis results, 
using the combined sample (main$+$expanded)
the correlations are more significant ($P =$1e-5, 4e-5, 0.102, 0.0057, $z$ from low to high),
again showing a flatter slope ($\alpha \sim$ 0.2, Figure~\ref{fig:lxsfr})
in each sub-sample than in the redshift-driven, overall correlation. 
The exception is the sub-sample with redshift range 1.0$< z < $1.5,
which shows a larger scatter. 
Our result agrees well with \citet{netzer09}, where an overall bi-variate 
linear correlation with similar slope was reported, 
and is consistent with \citet{stanley15}, 
in which using average IR luminosities,
including treatment of undetected sources, 
flat correlations were observed in similar redshift bins. 

The effect of different variability timescales for AGN and \sf\ activity has been studied
		on a quantitative base via hydrodynamic simulations
		 \citep[e.g.][]{gabornbournaud13, volonteri15a, thacker14, mcalpine17}
		 and through analytical models \citep[e.g.][]{hickox14}.
		 These studies have confirmed that different variability timescales 
		 can result in the observed
		flat trend in SFR as a function of AGN luminosity.
A flat slope does not rule out an underlying strong correlation,
given the large scatter introduced by the IR AGN/star formation degeneracy 
and the short time-scale AGN variability.
On the other hand, we do not see enhanced SFR towards higher AGN luminosity (\lx\ $> 10^{44.8}\ergs$),
as was reported in \citet{rosario12},
nor suppressed SFR towards luminous AGNs (the `feedback' effect).
Note that `feedback' can occur at different time scales 
\citep[e.g.][]{fabian12, gabornbournaud13, heckmannbest14, hopkins16},
	for instance, mechanical `feedback' from relativistic jets could occur in 0.1\,Myr scale,
	local `feedback' of the molecular cloud's dynamical evolution may take $\sim$1-10\,Myr, 
	while galaxy-wide `feedback', associated with gas cooling, may last much longer.

Based on our bivariate and PSRA results, 
as well as the correlation significance in the redshift limited sub-samples,
we conclude that there is
a weak (flat, $\alpha \sim$ 0.2) but significant
relation between SF and the luminosity of the central AGN.
After adding the expanded sample 
the correlations become more significant.
Since the correlation is positive, we see no evidence for the quenching of \sf\ 
by the AGN. However, we cannot exclude
the possibility that the feedback process either 
1) has significantly different timescales
	than the gas depletion time in these IR-bright AGNs (depends on the exact SFR
	and gas reservoir, thus could vary from a few Myr to a few 100 Myr),
or 2) introduces scatter that masks the underlying correlation.

\subsection{The Effect of Binning on the AGN-SF Correlation}
\label{sec:bin}
Given the rapid and significant variability of many AGNs, 
e.g. their flux can vary by 100 times in 0.1 Myr \mbox{\citep{keel12}},
it has been suggested that using an instantaneous X-ray luminosity could lead to large scatter
which smears out any intrinsic AGN-SF correlation \citep[e.g.][]{hickox14}.
Several studies have used the average X-ray or AGN luminosities and 
observed a positive correlation between AGN and SFR \citep{chen13, chen15, azadi15}.
Binning the data by different criteria
not only loses information from the dataset, 
but also projects any intrinsic \lagn-SFR relation onto a specific axis,
affecting the apparent slope of any correlation \citep{volonteri15b}.
We thus choose not to bin the data in our analysis (Section \ref{sec:correlation}, \ref{sec:psra}).
When stacking data below the detection limit, the average values
are biased by the outliers in each bin, 
especially when stacking a relatively small number of non-detections.
Instead, we defined an expanded sample which includes X-ray detected sources with `marginal' 
detections in the IR (See Section~\ref{sec:sample}), 
extending
below the flux limit to include fainter IR sources. This allows us to probe below the formal
flux limit while including as much information as possible from the expanded sample.
In our main sample, 
the average and median in each luminosity bin are
statistically identical based on the K-S test result (Table~\ref{tab:values}).

In Figure~\ref{fig:lxsfrcf}, we compare the best-fit correlations based on different binning methods
to test the variability scenario \citep{volonteri15a}.
In red is the average \lagn\ binned by \lirsf,
and in blue is the average \lirsf\ binned by \lagn.
Both show a positive correlation with $P < 0.000001$).
The average relation binned by \lagn\ (blue line, slope $k =0.59\,\pm\,$0.17)
agrees well with the correlation without binning (dot-dashed black line, $k =0.62\,\pm\,$0.05). 
Compared to the \lagn\ bins,
a strong correlation with a steeper slope is observed using \lirsf\ bins ($k = 1.11\,\pm$\,0.19). 
This effect is also seen in the \citet{azadi15} results (dark green stars in Fig.~\ref{fig:lxsfrcf}).

This slope change is a natural result of running a correlation test assuming a different dependent variable, 
but a similar change is also predicted by the 
more rapid variability of the AGN (Section \ref{sec:psra}), 
which explains
the lack of the AGN-SF correlation 
for samples that are mass- or SFR- selected \citep[e.g.][]{volonteri15b}.
Unfortunately, 
the degeneracy in both timescale and amplitude variation in the simulations,
as well as the large uncertainties in the observed luminosities,
make it difficult to predict on a quantitative basis
how much of the slope change is due to variability effect.
Future simulation work is needed to test the relative statistical
and variability contributions to the change in slope.

\subsection{The Ratio between SFR and BHAR}
\label{sec:ratio}
\begin{figure*} 
\begin{center}
\includegraphics[scale=0.7]{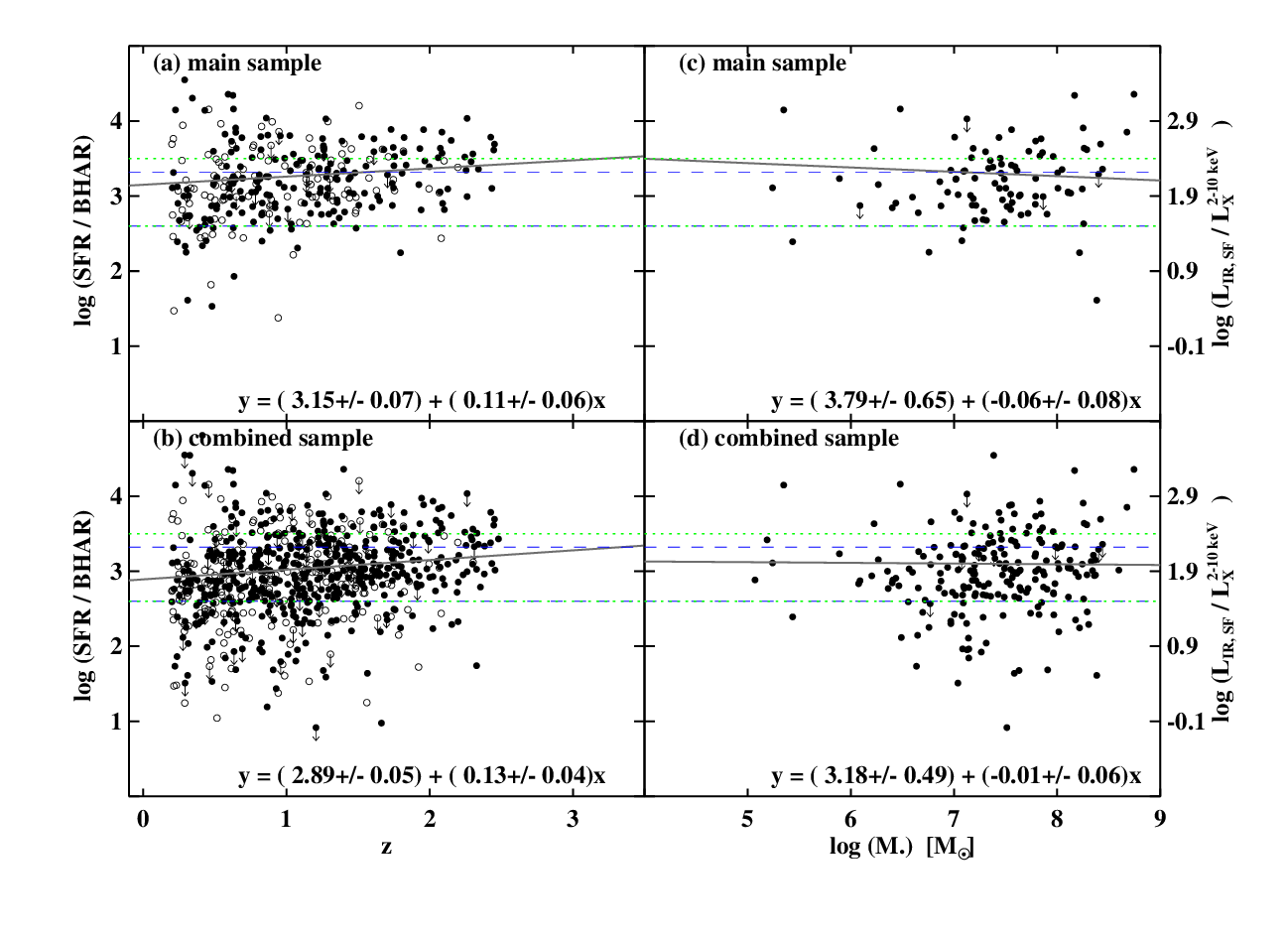}
\vspace{-20pt}
\end{center}
\caption{
The log (SFR/BHAR) ratios as a function of redshift (a \& b) and SMBH (c \& d) 
for the main sample and the combined sample (main$+$expanded).
Only type 1 AGNs with an $\mbh$ estimate are included in the right panels.
On the right hand y-axis we mark the values of corresponding \lirsf\ and hard X-ray luminosity \lx\ ratios.
Filled and open circles mark the type 1 (optical broad emission line or X-ray unobscured) and 
type 2 (optical narrow line or X-ray obscured) objects, respectively. 
The solid lines are the linear fits to the IR-bright AGN sample,
with the fitted function marked in each panel.
We observe a mean ratio 
of $\sim$ 3 for the mass formation/accretion ratios log (SFR/BHAR).
The intrinsic scatters (standard deviation) are 0.50 (a), 0.55 (b), 0.51(c) and 0.55 (d), respectively. 
Inclusion of the $z< 0.2$ and $z > 2.5$ supplementary sample yields consistent results (not plotted). 
The dashed blue lines mark the $\pm\,1\sigma$ range of the $M_*/\mbh$ ratios 
from \citet{mnh03}, 
and the dotted green lines mark the range in \citet{mnf01,mnd02}. 
\label{fig:ratio}}
\end{figure*}

To test if the AGN-SF relation evolves with redshift or $\mbh$, 
in this section we compare the relative strength of \sf\ and AGN,
represented by the ratio of \lirsf\ and \lx.
We find that the luminosity ratio does not evolve with redshift or $\mbh$.
Up to $z =$ 2.5, the main sample shows a non-zero ratio of:
\begin{equation} \label{eqn:ratio2}
{\rm log(SFR/BHAR)} = (3.15\,\pm\,0.07) + (0.11\pm0.06)z
\end{equation}
with a standard deviation of 0.50 (Figure~\ref{fig:ratio}, a).
We also compare the ratios for the subsample of broad-lined AGNs 
with a secure $\mbh$ estimate (Figure~\ref{fig:ratio}, c). 
Despite the incompleteness of low luminosity objects at high $z$,
SFR/BHAR shows an overall constant ratio across $\mbh$:
\begin{equation} \label{eqn:ratio22}
{\rm log(SFR/BHAR)} = (3.79\,\pm\,0.65) -(0.06\,\pm\,0.08) {\rm log(\mbh/\msun)}
\end{equation}
with a standard deviation of 0.52.
This confirms the correlation between AGN and star formation in the IR-bright AGN phase.

These results are consistent with the scenario 
in which both star formation and SMBH growth are dependent upon a common gas 
supply that feeds the AGN-host system. 
Candidate sources of the gas supply are: mergers, cosmic cold flows, 
or secular evolution \citep[e.g.][]{mullaney12b, dimatteo05, padovani17}.
Although we cannot distinguish the source of the gas reservoir,
the correlation indicates that for IR-active AGNs, 
both the core and the surrounding star formation---local/circumnuclear 
or on the galaxy scale---are growing in tandem on a cosmological basis. 
High resolution imaging is needed to trace the location of the star formation.
For our low-$z$ and high-$z$ subsamples,
the wide luminosity range, non-zero correlation slope, and large scatter 
indicate that there may be a mixture of major mergers and isolated galaxies.

Our results are also in general agreement with \citet[][M12]{mullaney12b},
despite the different sample selections.
The M12 sample consists of stacks from star-forming galaxies that
are both X-ray detected ($\sim$20\%) and undetected ($\sim$80\%).
Although the majority of the M12 sample have no X-ray detection,
a similar flat SFR/BHAR ratio was reported, 
with a constant value not evolving with redshift or stellar mass,
consistent with our results. 
Compared to M12 (207 X-ray detections $+$ 1037 X-ray non-detections), 
we have a larger sample with X-ray detections (323 $+$ 33 supplementary), 
plus the expanded sample of 558 AGNs below the formal IR flux limit.
At similar SFR, X-ray undetected targets have relatively 
lower BHAR and can drive the SFR/BHAR ratio higher, 
as confirmed by the higher mean ratio in M12, where log(SFR/BHAR) $=$ 3.2-3.5. 
On the other hand, at similar X-ray luminosity,
the inclusion of sources that are not detected in the IR would 
lower the SFR/BHAR ratio, 
as demonstrated later with our combined sample. 
Regardless of the sample used,
this growth ratio of log(SFR/BHAR) $\sim$ 3 is constant over redshift,
and dominates the final mass ratios if integrated over a long enough accretion history,
thus leading to the locally observed mass ratios at 
log ($M_{*}/\mbh$) $\sim$ 3 \citep[e.g.][see also Equation (7) of M12]{mnf01,mnd02,mnh03}.

A similar result of constant SFR/BHAR ratios, independent of redshift,
was reported in \citet{silverman09},
though with a 10 times lower log(SFR/BHAR) ratio at $\sim$2.
Since the sample selection (X-ray luminosity based) and SFR estimates ($[OII]$ emission line based)
are distinctively different from this work,
it is difficult to compare the results directly.
Constant SFR/BHAR ratios have also been predicted by simulations,
where the galaxy and BH can be modulated by torque-limited growth 
along the bulge-BH mass plane from $z=$ 4 to $z=$ 0 \citep[e.g.][]{anglesalczar15}, 
though large scatter and distinct variations are often noticed \citep[e.g.][]{thacker14, mcalpine17}.

To check for Malmquist bias, 
we also fit the SFR/BHAR $vs$ $z$ and $\mbh$ for
the main$+$expanded combined sample (Figure~\ref{fig:ratio}, b,d).
The resulting average ratios in log scale are (2.89 $\pm$ 0.05) for SFR/BHAR $vs$ $z$, 
and (3.18 $\pm$ 0.49) for SFR/BHAR $vs$ $\mbh$.
Using the expanded sample alone results in similar log ratios:
(2.89 $\pm$ 0.05) for SFR/BHAR $vs$ $z$, 
and (3.15 $\pm$ 0.07) for SFR/BHAR $vs$ $\mbh$,
consistent with the main sample results within errors. 
Both relations have a standard deviation of 0.55.
We note that at a given \lirsf/\lx\ or SFR/BHAR ratio,
the AGN contribution to the total IR luminosity (f$=L_{\rm IR, AGN}$/\lir, see Figure~\ref{fig:agnf} and Sec.~\ref{sec:sed})
ranges from insignificant ($f_{\rm AGN}<$ 0.2) to dominant ($f_{\rm AGN}>$ 0.5).

The mean SFR/BHAR ratio 
is consistent with the ratio of stellar to SMBH mass seen in local galaxies.
The BH-host bulge mass relation is well-established at low $z$ for AGNs:
log ($M_{\rm bulge}/\mbh$) $=$ 2.81 $\pm$ 0.36 \citep{mnh03}, 
log ($M_{\rm bulge}/\mbh$) $=$ 2.90 $\pm$ 0.45 \citep{mnf01,mnd02}.  
Assuming the log ($M_{\rm bulge}/M_{\rm *,total}$) ratio is 
around $-0.15$ \citep[][median for 660,000 SDSS DR7 galaxies]{mendel14},  
we convert the BH-bulge relation
to 
log ($M_{*}/\mbh$) $=$ (2.96 $\pm$ 0.36) and log ($M_{*}/\mbh$) $=$ (3.05 $\pm$ 0.45) (blue and green dashed lines 
in Figure~\ref{fig:ratio}),
both consistent with Equation~\ref{eqn:ratio2}.
Although IR-detected and undetected AGNs are reported 
to have similar host galaxy stellar masses \citep[e.g.][]{santini12, rovilos12, rosario13a},
significantly different log ($M_{*}/\mbh$) ratios 
from those quoted above have also been 
reported (e.g. 2.6 $\pm$ 0.4 in \citet{knh13},
and 3.6 $\pm$ 0.5 \citep{rnv15}).
The difference in the accumulated mass ratios is sample dependent,
and varies, from high to low, 
across quiescent, bulgeless/pseudobulges,
classical giant elliptical and bulge galaxies \citep{knh13}.

Compared to the literature values for 
log $(M_{*}/\mbh)$ (e.g. blue and green lines in Figure~\ref{fig:ratio}), 
our sample shows a larger scatter (0.50 $vs$ 0.36 in \citet{mnh03}, $vs$ 0.45 in \citet{mnf01,mnd02}, all in log scale). 
This is due to one or more of the following factors:
(a). the uncertainties in the SFR and BHAR estimates; 
(b). the large range of the instantaneous ratios due to rapid AGN variability;
(c). the unknown host galaxy morphology---our sample is not restricted to bulges or ellipticals,
and could include spirals and other type of galaxies 
that may not share the same mass ratios. 
However the mean SFR/BHAR ratio
is not related to mass or redshift (Figure~\ref{fig:ratio}),
indicating that for objects with currently active growth of black hole and star formation,
the central BH and the stellar mass grow at a similar rate on average. 
The large scatter in the mass ratios is likely the result of
the different evolutionary paths and stages of the AGN systems in the sample.

\section{SUMMARY}
\label{sec:summary}
We have constructed a sample of 323 IR-bright AGNs
with \lx\, $> 10^{42} \ergs$ at 0.2 $< z <$ 2.5
in the $\sim$ 11\, deg$^2$ XMM-LSS field (the main sample). 
All targets are detected in both hard X-ray and FIR with 
either spectroscopic or photometric redshift measurements.
The majority of the sample (65\%) are type 1 objects,
and 86\% have dust mass greater than $10^8\,\msun$.
This IR-bright AGN sample is thus dominated by type 1 AGNs with 
significant dust and star formation in the AGN-host system.
For comparison,
and to expand the luminosity parameter space,
we also construct 
an expanded sample of 558, X-ray AGNs with 1-3$\sigma$ IR detections, formally undetected; 
and a supplementary sample of 33, $z<0.2$ or $z > 2.5$ IR-bright AGNs.
Our main results are summarized as follows:

\begin{enumerate}
\item 
We find significant bivariate AGN-IR correlations 
between the absorption-corrected X-ray luminosity and 
total IR luminosities for the IR-bright AGN main sample, 
including integrated luminosities \lir\ (8-1000\micron), \lfir\ (30-1000\micron), 
and at individual wavelengths ($L_{\rm 60}$, $L_{\rm 100}$).
The \lirsf\ and \lagn\ correlation 
has a power law slope of $(0.62\,\pm\,0.05)$,
and a probability of $P <$\,0.0001.  
This slope does not vary significantly towards fainter flux-limits.

\item 
We find that AGN can contribute significantly to the IR and FIR fluxes and luminosities,
and this contribution ranges from negligible up to almost 100\%. 
On average, the AGN contributions to the total IR luminosity (8-1000\micron) are
11\% for the main sample,
and 23\% for the expanded sample. 
The total ${\rm L_{FIR}^{300-1000}}$ is consistent with 
the AGN-removed ${\rm L_{IR, SF}}$ within the errors for $>$92\% of the main sample,   
and can be used as a proxy for ${\rm L_{IR, SF}}$.
We find that single-band luminosity at longer wavelengths ($L_{\rm 60}$ and $L_{\rm 100}$)
suffers the least AGN contamination ($\sim$2\%).
 
\item 
The application of a partial correlation test (PSRA) to determine the dominant variable 
for the main sample leads 
to the conclusion that the bivariate AGN-IR correlations 
are primarily driven by redshift (Table~\ref{tab:pca}). 

\item
Binning the data by either IR or X-ray luminosity affects the observed correlation significance and slopes.

\item
Using $L_{\rm 60}$ and $L_{\rm 100}$, as proxies for star formation,
PSRA tests reveal a strong residual AGN-SF correlation with a slope of $\sim$0.2 and high significance ($P < $0.005),
beyond that resulting from the redshift
in our combined sample (main$+$expanded).
The lack of a correlation when using \lirsf\ is likely 
a result of the large uncertainties in the AGN-subtraction.

\item
Significant, flatter ($\alpha  \sim$ 0.2) \lagn-\lirsf\ correlations are found in 3 of 4 sub-samples 
covering a smaller redshift range (so as to reduce the redshift effect). 
The correlation significance increases when the expanded sample is also included. 
However the large scatter and small sub-samples result in large errors on the derived slopes.
A larger IR-detected sample with a wider IR luminosity range 
at each redshift will be valuable to confirm these results.
While we see no evidence for star formation being quenched by AGN activity, 
it remains possible that this occurs on significantly different timescales than 
probed by our study.

\item
There is no evidence that the relationship between 
AGN luminosity and IR luminosity 
changes with black hole mass or Eddington ratios.

\item
The average ratio of the star formation and BH accretion rate is:
 log (SFR/BHAR) $\sim$ 3.15, with a deviation of 0.50 
 ($\sim$2.89 $\pm$ 0.55 for the combined sample), 
 independent of redshift or SMBH mass, but with a wide dispersion.
The average SFR/BHAR ratio is consistent with the mean observed
$M_*/\mbh$ ratio found in local galaxies.
The consistent averages support a scenario in which 
a SMBH and its host galaxy both grow from
a common gas supply, when averaged over long time periods.
Unlike earlier results that suggested two formation paths \citep[e.g.][]{lutz10, shao10, rosario12, santini12},
our overall correlation combined with the similar results in different
redshift bins,
suggest that it is not likely that AGN and \sf\ are completely
unrelated, 
nor that they are dominated by feedback in which an AGN quenches the \sf.
\end{enumerate}

\section*{Acknowledgements}
We thank the anonymous referee for constructive and detailed comments which
improved and clarified this paper.

YSD would like to thank Lee Armus, Nick Scoville, Phillip Hopkins, Chris Hayward,
Xiangcheng Ma, and Ranga R. Chary for helpful discussions. 
BJW gratefully acknowledged the support of NASA Contract NAS8-03060
(CXC).

This work is sponsored in part by the Chinese Academy of Sciences (CAS), through a grant to
the CAS South America Center for Astronomy (CASSACA) based in Santiago, Chile.
This research has made use of data from HerMES project (http://hermes.sussex.ac.uk/). 
HerMES is a Herschel Key Programme utilising Guaranteed Time from 
the SPIRE instrument team, ESAC scientists and a mission scientist.
The HerMES data was accessed through the Herschel Database in Marseille (HeDaM - http://hedam.lam.fr) 
operated by CeSAM and hosted by the Laboratoire d'Astrophysique de Marseille.

\begin{table*}
\begin{center}
\caption{The range of X-ray and IR luminosities in bins of redshift}
\begin{tabular}{lrrrrrrr}
\hline
\hline
Redshift bins   & $ z\leq$ 0.2  & 0.2 $< z \leq$ 0.5 & 0.5 $< z \leq $ 1.0 & 1.0 $< z \leq$ 1.5 & 1.5 $< z \leq $ 2.5  &  $z > 2.5$ & Total \\   
\hline
Main Sample  & ... & 63 (22\%) & 106 (17\%) & 88 (14\%) & 66 (11\%) & ...  & 323 (14\%)\\
expanded sample  & ... & 76 (27\%) & 165 (26\%) & 156 (24\%) & 161 (27\%) & ...  & 558 (23\%)\\
Supplementary Sample  & 20(16\%)  & ... & ... & ...  & ... & 12(10\%) & 32 (2\%)\\
\hline

\hline
\hline
\label{tab:sample}
\end{tabular}
\end{center} 
\raggedright
Note: In parenthesis are the percentage among the parent sample of 
2,399 hard X-ray detected AGNs in the same redshift range. 
\end{table*}

\begin{landscape}

\begin{table}
\begin{center}
\caption{Derived properties for IR-bright AGNs}
\begin{tabular}{lccccccccccccc}
\hline
\hline
 Xcatname         &          redshift  &  zflag   &  $T_{\rm dust}$     &  $\alpha$   & HR     &    $\rm N_{H}$ (int)  &    \lx\  &  $L_{\rm IR, AGN}$  & \lirsf\    & SFR  &  log$M_{\rm dust}$ &  log$\mbh$  & flag \\
 & & & (K) & & & (cm$^{-2}$) & ($\ergs$) &($\ergs$) & ($\ergs$) & ($\msun\,yr^{-1}$) & ($\msun$) & ($\msun$)  & \\
 (1) & (2) & (3) & (4) & (5) & (6) & (7) & (8) & (9)& (10)& (11) & (12) & (13)& (14) \\
\hline
2XLSSd J021324.6-033512 &  1.142 &  1  &  47.6$\pm$7.3 &2.0$\pm$0.0  &  -0.06 & 2.8e+22 & 44.64  & 45.9$\pm$0.7 & 46.7$\pm$1.0  & 2110 &  8.5$\pm$1.3 & 9.4 & 1 \\
2XLSSd J021407.8-035309 &  0.987 &  1  &  18.7$\pm$1.7 &0.4$\pm$0.0  &   0.28 & 5.8e+22 & 44.13  & 45.1$\pm$0.6 & 46.4$\pm$1.0  & 1100 &  9.2$\pm$1.1 & 8.5 & 1 \\
2XLSSd J021418.7-033934 &  1.136 &  2  &  20.4$\pm$0.9 &1.4$\pm$0.0  &  -0.08 & 2.6e+22 & 44.24  & 45.3$\pm$0.6 & 46.2$\pm$1.0  & 730  &  9.4$\pm$1.3 & ... & 1 \\
2XLSSd J021434.2-035553 &  1.426 &  1  &  51.9$\pm$1.3 &3.0$\pm$0.1  &  -0.48 & 7.7e+20 & 45.14  & 46.8$\pm$0.8 & 0  & 0      &  8.3$\pm$0.9 & 8.6 & 1 \\
2XLSSd J021451.6-035339 &  0.614 &  1  &  29.1$\pm$5.2 &2.4$\pm$0.2  &  -0.52 & 0.0     & 43.18  & 43.8$\pm$0.5 & 45.6$\pm$0.9  & 160  &  8.3$\pm$1.2 & ... & 1 \\
\hline
\hline
\label{tab:data}
\end{tabular}
\end{center} 
\raggedright
Notes: 
(1) Object identification same as in \citet{chiappetti13}. 
(2) Redshift of the object.
(3) Redshift flag, 1 for spec-$z$, 2 for photo-$z$.
(4) \& (5) Peak dust temperature and power-law index derived from SED fitting as described in Sec.~\ref{sec:sed}.
(6) X-ray Hardness ratio (HR$=$(H $-$ S)/(H $+$ S) ) based on net count rates. 
(7) Intrinsic column density derived from redshift and HR as described in Sec.~\ref{sec:sample}. 
If the object is only detected in the hard X-ray, a lower limit of 1.00e$+$23$+$ is assigned. 
If the object has an HR $< -0.5$, no $N_{\rm H}$ correction was made and a value of `0' was assigned.
(8) Obscuration corrected rest-frame X-ray luminosity  (2-10\,keV).
(9) AGN contributed infrared luminosity (8-1000\,$\mu$m), based on X-ray luminosity converted 6$\mu$m luminosity \citep{stern15} and SED template from \citep{dai12}. 
(10) AGN-subtracted infrared luminosity (8-1000\,$\mu$m). A value of `0' marks purely AGN driven IR luminosity.  
(11) SFR derived from (10) using the \citet{kennicutt98} relation. A value of `0' marks purely AGN driven IR luminosity.  
(12) Dust mass derived from FIR photometry as described in Sec.~\ref{sec:sed}.
(13) SMBH mass derived for the subsample with optical broad emission lines as described in Sec.~\ref{sec:er}.
(14) Sample flag, 1 for main sample, 2 for expanded sample, and 3 for supplementary sample. 
This table is available in its entirety with a machine-readable form in the online journal.
\end{table}

\clearpage
\end{landscape}

\begin{table*}%[htdp]
\begin{center}
\caption{Bivariate Correlation Analysis for the Main Sample}
\begin{tabular}{lclc}
\hline
\hline
 Correlation & Probability &  Correlation & Probability\\ [1pt]
\hline
\hline
$L_{\rm X}$ vs $L_{\rm IR, SF}$ & $<0.0001$ & $L_{\rm X}$ vs $L_{\rm FIR, SF}$ &  $<0.0001$ \\ [1pt]
$L_{\rm X}$ vs $L_{\rm IR}$ & $< 0.0001$ & $L_{\rm X}$ vs $L_{\rm FIR}$ & $< 0.0001$ \\ [1pt]
$L_{\rm X}$ vs $L_{60}$ & $< 0.0001$ & $L_{\rm X}$ vs $L_{100}$ &  $<$0.0001 \\ [1pt]
$F_{\rm X}$ vs $F_{60}$ & 0.5980 & $F_{\rm X}$ vs $F_{100}$ & 0.2324\\ [1pt]
\hline
\hline
\label{tab:bcamain}
\end{tabular}
\end{center} 
\end{table*}

\begin{table*}%[htdp]
\begin{center}
\caption{Bivariate Correlation Analysis for the Combined (main$+$expanded$+$supplementary) Sample}
\begin{tabular}{lclc}
\hline
\hline
 Correlation & Probability &  Correlation & Probability\\ [1pt]
\hline
\hline
$L_{\rm X}$ vs $L_{\rm IR, SF}$ & $<0.0001$ & $L_{\rm X}$ vs $L_{\rm FIR, SF}$ &  $<0.0001$ \\ [1pt]
$L_{\rm X}$ vs $L_{\rm IR}$ & $< 0.0001$ & $L_{\rm X}$ vs $L_{\rm FIR}$ & $< 0.0001$ \\ [1pt]
$L_{\rm X}$ vs $L_{60}$ & $< 0.0001$ & $L_{\rm X}$ vs $L_{100}$ &  $<$0.0001 \\ [1pt]
$F_{\rm X}$ vs $F_{60}$ & 0.0106 & $F_{\rm X}$ vs $F_{100}$ & 0.0073 \\ [1pt]
\hline
\hline
\label{tab:bca}
\end{tabular}
\end{center} 
\end{table*}

%\tablenum{5}
\begin{table}
\begin{center}
\caption{Partial Correlation Analysis}
\begin{tabular}{|l|rc|rc|}
\hline\hline
\multicolumn{1}{|c}{Correlation} &
\multicolumn{2}{|c}{Main Sample} &
\multicolumn{2}{|c|}{Combined Sample} \\
\multicolumn{1}{|c}{}  &
\multicolumn{1}{|c}{$P$} &
\multicolumn{1}{c|}{$r$} &
\multicolumn{1}{c}{$P$} &
\multicolumn{1}{c|}{$r$} \\ \hline
$L_X$ vs $L_{60}$        &    0.377& 0.018&$<$ 0.005& 0.218\\
$F_X$ vs $F_{60}$        &    0.382& -0.017&$<$ 0.005& 0.267\\
&&&&\\
$L_X$ vs $L_{100}$       &    0.388& 0.016&$<$ 0.005& 0.152\\
$F_X$ vs $F_{100}$       & $>$ 0.400& 0.011&$<$ 0.005& 0.205\\
&&&&\\
$L_X$ vs $L_{IR}$       &    $<$ 0.005 & 0.163& 0.011& 0.129\\
$L_X$ vs $L_{FIR}$    &    0.032& 0.105&    0.061& 0.088 \\
&&&&\\
$L_{\rm AGN}$ vs $L_{\rm IR,SF}$ & 0.297 & 0.030  &  $>$0.400 &  -0.014\\
$L_{\rm AGN}$ vs $L_{\rm FIR,SF}$	& 0.087 & 0.070  &  $>$0.400 & -0.007\\
\hline
\label{tab:pca}
\end{tabular}
\raggedright
\\
Notes: 
$P$ is the partial Spearman rank probability and $r$
the partial correlation coefficient for a correlation between
the listed parameters 
occurring by chance, given that both variables depend on
redshift. Partial correlations between all luminosities and
redshift are universally significant ($P<0.005$).
Combined sample includes the main, expanded, and supplementary samples.
\end{center}
\end{table}

\begin{table*}
\begin{center}
\caption{Derived average properties in each luminosity bin for the main sample}
\begin{tabular}{ccccccc}
\hline
luminosity range             & $N_{det}$     & $z$ range  &  $<log$\lagn$>$  & $<log$\lirsf$>$ & $<logM_{\rm BH}>$  & $<ER>$  \\
(1) &(2) &(3) &(4)&(5)&(6) &(7)\\
\hline
 \lagn\ bins &&&&&&\\ 
43.0-43.5   &   13   &   0.200-0.430   & 43.33$^{+0.30}_{-0.17}$ &   44.96$^{+0.69}_{-1.39}$  &    6.33$^{+0.09}_{-0.06}$  &  0.06$^{+0.03}_{-0.02}$ \\ 
43.5-44.0   &   17   &   0.208-0.593   & 43.80$^{+0.28}_{-0.17}$ &   44.80$^{+0.82}_{-1.19}$  &    9.75 &  ...  \\ 
44.0-44.5   &   51   &   0.205-0.900   & 44.28$^{+0.28}_{-0.21}$ &   45.24$^{+1.77}_{-1.25}$  &    7.43$^{+1.26}_{-1.74}$  &  0.29$^{+0.29}_{-0.65}$ \\ 
44.5-45.0   &   49   &   0.238-1.730   & 44.76$^{+0.26}_{-0.24}$ &   45.66$^{+0.87}_{-0.79}$  &    8.10$^{+1.01}_{-1.15}$  &  0.07$^{+0.07}_{-0.27}$ \\ 
45.0-45.5   &   66   &   0.310-2.261   & 45.25$^{+0.25}_{-0.25}$ &   45.89$^{+1.73}_{-1.24}$  &    8.25$^{+0.85}_{-1.13}$  &  0.12$^{+0.11}_{-0.41}$ \\ 
45.5-46.0   &   71   &   0.316-2.447   & 45.75$^{+0.24}_{-0.25}$ &   46.24$^{+2.97}_{-0.76}$  &    8.58$^{+0.58}_{-0.52}$  &  0.17$^{+0.15}_{-0.49}$ \\ 
46.0-46.5   &   37   &   0.776-2.301   & 46.18$^{+0.17}_{-0.29}$ &   46.51$^{+1.90}_{-0.68}$  &    8.97$^{+0.57}_{-0.71}$  &  0.17$^{+0.15}_{-0.22}$ \\ 
46.5-47.0   &   14   &   1.077-2.452   & 46.65$^{+0.14}_{-0.20}$ &   46.65$^{+0.91}_{-0.70}$  &    9.13$^{+0.58}_{-0.31}$  &  0.32$^{+0.21}_{-0.49}$ \\ 
\hline
\lirsf\ bins &&&&&&\\
44.0-44.5   &  14 &  0.203-0.457 &  43.81$^{+0.55}_{-1.47}$  &  44.34$^{+0.23}_{-0.15}$  &    7.00$^{+0.76}_{-0.76}$  &     0.05$^{+0.03}_{-0.03}$  \\
44.5-44.0   &  30 &  0.200-0.793 &  44.36$^{+1.20}_{-1.71}$  &  44.78$^{+0.27}_{-0.21}$  &    7.43$^{+1.26}_{-1.95}$  &     0.41$^{+0.40}_{-0.53}$  \\
45.0-45.5   &  55 &  0.225-1.034 &  44.66$^{+1.39}_{-1.37}$  &  45.27$^{+0.27}_{-0.23}$  &    7.67$^{+1.32}_{-0.85}$  &     0.11$^{+0.10}_{-0.23}$  \\
45.5-45.0   &  65 &  0.343-1.482 &  44.95$^{+1.58}_{-1.56}$  &  45.77$^{+0.27}_{-0.22}$  &    8.27$^{+0.87}_{-1.47}$  &     0.13$^{+0.13}_{-0.39}$  \\
46.0-46.5   &  95 &  0.249-1.842 &  45.48$^{+2.45}_{-1.54}$  &  46.26$^{+0.25}_{-0.23}$  &    8.51$^{+2.12}_{-0.75}$  &     0.12$^{+0.12}_{-0.26}$  \\
46.5-46.0   &  49 &  1.078-2.447 &  46.06$^{+0.65}_{-0.96}$  &  46.74$^{+0.23}_{-0.26}$  &    8.80$^{+0.79}_{-0.60}$  &     0.23$^{+0.19}_{-0.57}$  \\
47.0-47.5   &  10 &  1.507-2.452 &  46.32$^{+0.83}_{-0.53}$  &  47.13$^{+0.09}_{-0.22}$  &    9.27$^{+0.71}_{-0.41}$   &    0.12$^{+0.10}_{-0.13}$  \\
\hline
\label{tab:values}
\end{tabular}
\end{center} 
\raggedright
Notes: 
(1)Luminosity range in \ergs, 
(2) number of IR-bright AGNs in the selected bin, excluding SFR$=$0 objects, 
(3) redshift range for objects in the bin,
(4) average AGN bolometric luminosity in \ergs, (5) average AGN-removed IR luminosity in \ergs, 
(6) average SMBH mass in $\msun$ for the subsample in the bin with a mass estimate,
(7) average Eddington ratio for the subsample in the bin with a mass estimate.
The errors in column (4)-(7) indicate the range for the binned objects. 
\end{table*}

%%%%%%%%%%%%%%%%%%%%%%%%%%%%%%%%%%%%%%%%%%%%%%%%%%



%\bibliographystyle{mnras}


%%%%%%%%%%%%%%%%%%%%%%%%%%%%%%%%%%%%%%%%%%%%%%%%%%

%%%%%%%%%%%%%%%%% APPENDICES %%%%%%%%%%%%%%%%%%%%%

%%%%%%%%%%%%%%%%%%%%%%%%%%%%%%%%%%%%%%%%%%%%%%%%%%

% Don't change these lines
\bsp	% typesetting comment
\label{lastpage}
\end{CJK}
\end{document}